\documentclass[journal]{IEEEtran}
\def\mode{3} 
\def\redactionmode{3}
\def\alttextmode{0} 
\usepackage{adjustbox}              
\usepackage{amsmath,amssymb}
\usepackage[english]{babel}
\usepackage{cite}					
\usepackage[dvipsnames]{xcolor} 
\usepackage[oldvoltagedirection]{circuitikz}
\usetikzlibrary{patterns}
\usepackage[hidelinks]{hyperref} 
\usepackage{cleveref}
\usepackage{comment}  				

\usepackage{enumitem}				
\usepackage[T1]{fontenc}
\usepackage{graphicx}
	\graphicspath{{./img/}}		
\usepackage[utf8]{inputenc}
\usepackage{multirow}				
\usepackage{pgfplots}
\pgfplotsset{compat=1.16}
\usepackage{tabularx} 				
\usepackage{threeparttable}
\usepackage{subfigure}
\usepackage{soul} 					
\usepackage{xargs}    				

\newcommand{\red}[1]{#1}
\newcommand{\green}[1]{#1}
\newcommand{\blue}[1]{#1}

\newcounter{documentmode}
\includecomment{conf}  	
\newcommand{\setdocmode}{%
  \ifcase\number\value{documentmode}
    \renewcommand{\red}[1]{\color{red}##1\color{black}{}}
    \renewcommand{\blue}[1]{\color{blue}##1\color{black}{}}
    \renewcommand{\green}[1]{\color{olive}##1\color{black}{}}
  \or
    \usepackage{draftwatermark}
	\SetWatermarkScale{1.35}
	\SetWatermarkAngle{56}
	\SetWatermarkLightness{0.9}
	\SetWatermarkHorCenter{0.9\textwidth}
	\SetWatermarkVerCenter{0.7\textheight}
  \or
	\usepackage{draftwatermark}
	\SetWatermarkText{CONFIDENTIAL}
	\SetWatermarkScale{0.67}
	\SetWatermarkAngle{56}
	\SetWatermarkLightness{0.9}
	\SetWatermarkHorCenter{0.6\textwidth}
	\SetWatermarkVerCenter{0.6\textheight}
  \else
    \excludecomment{conf}
  \fi
}

\if\mode0
\usepackage[prependcaption, textsize=footnotesize]{todonotes}
\else
\usepackage[disable, prependcaption, textsize=footnotesize]{todonotes}

\fi

\setdocmode

\newcounter{redmode}
\includecomment{2Detailed}			
\includecomment{Uncritical}			
\includecomment{2SaveSpace}			
\includecomment{IfMustDelete}			
\newcommand{\redact}{%
  \ifcase\number\value{redmode}
    \includecomment{2Detailed}
    \includecomment{Uncritical}
    \includecomment{2SaveSpace}
    \includecomment{IfMustDelete}
  \or
    \excludecomment{2Detailed}
    \includecomment{Uncritical}
    \includecomment{2SaveSpace}
    \includecomment{IfMustDelete}
  \or
    \excludecomment{2Detailed}
    \excludecomment{Uncritical}
    \includecomment{2SaveSpace}
    \includecomment{IfMustDelete}
  \or
    \excludecomment{2Detailed}
    \excludecomment{Uncritical}
    \excludecomment{2SaveSpace}
    \includecomment{IfMustDelete}
  \or
    \excludecomment{2Detailed}
    \excludecomment{Uncritical}
    \excludecomment{2SaveSpace}
    \excludecomment{IfMustDelete}
  \fi
}
\redact

\newcounter{alttxtmode}
\includecomment{TextA}			
\includecomment{TextB}			
\newcommand{\alttext}{%
  \ifcase\number\value{alttxtmode}
    \includecomment{TextA}
    \includecomment{TextB}
  \or
    \excludecomment{TextA}
    \includecomment{TextB}
  \or
    \includecomment{TextA}
    \excludecomment{TextB}
  \or
    \excludecomment{TextA}
    \excludecomment{TextB}
  \fi
}
\alttext

\usepackage[acronym]{glossaries} 

\newacronym{asd}{ASD}{Autism Spectrum Disorder}
\newacronym{cart}{CART}{Classification And Regression Tree}
\newacronym{dfa}{DFA}{Discriminant Function Analysis} 
\newacronym{ecg}{ECG}{Electrocardiography}
\newacronym{skin_eda}{EDA}{Electrodermal activity} 		
\newacronym{ews}{EWS}{Early Warning Score}
\newacronym{ga}{GA}{Genetic Algorithm}
\newacronym{hr}{HR}{Heart Rate}
\newacronym{ldf}{LDF}{Linear Discriminant Function}
\newacronym{rmsdd}{RMSDD}{Root-Mean Square of Successive Differences}
\newacronym{rsvm}{RSVM}{Reputation-driven Support Vector Machine }
\newacronym{saews}{SA-EWS}{Self-Aware Early Warning Score}
\newacronym{sam}{SAM}{Self-Assessment Manikin} 
\newacronym{selphys}{SelPhyS}{Self-aware cyber-Physical System}
\newacronym{sdnn}{SDNN}{Standard Deviation Normal-to-Normal-Intervals}
\newacronym{skt}{SKT}{Skin Temperature}
\newacronym{som}{SOM}{Self-Organizing Map}
\newacronym{whs}{WHS}{Wearable Health-care Systems}

\newacronym{ai}{AI}{Artificial Intelligence}
\newacronym{ann}{ANN}{Artificial Neural Network}
\newacronym{bpn}{BPN}{Back-Propagation Neural Network}
\newacronym{bpnn}{BPNN}{back-propagation neural network}
\newacronym{cnn}{CNN}{Convolutional Neural Network}
\newacronym{flop}{FLOP}{Floating Point Operation}
\newacronym{icnn}{ICNN}{Iterative Convolutional Neural Network}
\newacronym{svm}{SVM}{Support Vector Machine}
\newacronym{nn}{NN}{Neural Network}
\newacronym{nb}{NB}{Naive Bayesian}
\newacronym{mcs}{MCS}{Multiple Classifier System}
\newacronym{ml}{ML}{Machine Learning}
\newacronym{ucnn}{$\mu$CNN}{Micro CNN}

\newacronym{brs}{BRS}{Bipolar Resistive Switch-based logic}
\newacronym{cnf}{CNF}{Conjunctive Normal Form}
\newacronym{crs}{CRS}{Complementary Resistive Switch-based logic}
\newacronym{dnf}{DNF}{Disjunctive Normal Form}
\newacronym{ecm}{ECM}{Electrochemical Metallization}
\newacronym{fpm}{FPM}{Forward Polarized Memristor}
\newacronym{hfo}{$HfO_x$}{Hafnium Oxide}
\newacronym{hrs}{HRS}{High Resistance State}
\newacronym{imc}{IMC}{In-Memory Computation}
\newacronym{imply}{IMPLY}{Material Implication}
\newacronym{imp}{IMP}{In-Memory Processing}
\newacronym{lim}{LIM}{Logic in Memory}
\newacronym{lrs}{LRS}{Low Resistance State}
\newacronym{ltg}{LTG}{Logic Threshold Gate}
\newacronym{magic}{MAGIC}{Memristor-Aided Logic}
\newacronym{mecoins}{Me-Coin}{Memristor-based Computation In-memory}
\newacronym{nimp}{NIMP}{NOT IMPLY}
\newacronym{pc}{PC}{Phase Change}
\newacronym{pcm}{PCM}{Phase Change Memory}
\newacronym{pim}{PIM}{Processing in Memory}
\newacronym{reram}{ReRAM}{Resistive Random Access Memory}
\newacronym{rpm}{RPM}{Reversely Polarized Memristor}
\newacronym{stt}{STT}{Spin Transfer Torque}
\newacronym{tao}{$TaO_x$}{Tantalum Oxide}
\newacronym{tio}{$TiO_2$}{Titanium dioxide}
\newacronym{vc}{VC}{Valence Change}

\newacronym{aco}{ACO}{Autonomous Cooperating Object}
\newacronym{afdd}{AFDD}{Automated Fault Detection and Diagnostic}
\newacronym{ca}{CA}{Continuous Average}
\newacronym{cah}{CAH}{Context-Aware Health Monitoring}
\newacronym{cam}{CCAM}{Confidence-based Context-Aware condition Monitoring}
\newacronym[plural=DABs,longplural={Discrete Average Blocks}]{dab}{DAB}{Discrete Average Block}
\newacronym[plural=KPNs,longplural={Kahn Process Networks}]{kpn}{KPN}{Kahn Process Networks} 
\newacronym{mape-k}{MAPE-K}{Monitor-Analyze-Plan-Execute over a shared Knowledge}
\newacronym[plural=MoCs,longplural={Models of Computation}]{moc}{MoC}{Model of Computation}
\newacronym{oda}{ODA}{Observe-Decide-Act}
\newacronym{pca}{PCA}{Principal Component Analysis}
\newacronym{rosa}{RoSA}{Research on Self-Awareness}
\newacronym{sa}{SA}{Self-Aware}
\newacronym{saness}{SA}{Self-Awareness}
\newacronym{sahm}{SAHM}{Self-Aware Health Monitoring}
\newacronym{samba}{SAMBA}{Self-Aware health Monitoring and Bio-inspired coordination for distributed Automation systems}
\newacronym{sh}{SH}{State Handler}

\newacronym{cps}{CPS}{Cyber-Physical System}
\newacronym{cpps}{CPPS}{Cyber-Physical Production System}
\newacronym{dsr}{DSR}{Down-Sampling Rate}
\newacronym{dum}{DuM}{Device under Monitoring}
\newacronym{es}{ES}{Embedded System}
\newacronym{mes}{MES}{Manufacturing Execution System}
\newacronym{sos}{SoS}{System of Systems}
\newacronym{suo}{SuO}{System under Observation}

\newacronym{abi}{ABI}{Application Binary Interface}
\newacronym{adc}{ADC}{Analog-to-Digital Converter}
\newacronym{aes}{AES}{Advanced Encryption Standard}
\newacronym{alu}{ALU}{Arithmetic Logic Unit}
\newacronym{api}{API}{Application Programming Interface}
\newacronym{asic}{ASIC}{Application Specific Integrated Circuit}
\newacronym{asoc}{ASOC}{Autonomic System-on-Chip platform}
\newacronym{axi}{AXI}{Advanced eXtensible Interface Bus}
\newacronym{bram}{BRAM}{Block Random Access Memory}
\newacronym{cdt}{CDT}{C/C++ Development Tooling}
\newacronym{clb}{CLB}{Configuarable Logic Block}
\newacronym{cmos}{CMOS}{Complementary Metal-Oxide Semiconductor}
\newacronym{cp}{CP}{Clock Pulse}
\newacronym{cpi}{CPI}{Cycles Per Instruction}
\newacronym{cpu}{CPU}{Central Processing Unit} 
\newacronym{cpsoc}{CPSoC}{Cyber-Physical System-on-Chip}
\newacronym{cu}{CU}{Compute Unit}
\newacronym{cuda}{CUDA}{Compute Unified Device Architecture}
\newacronym{dac}{DAC}{Digital to Analog Converter}
\newacronym{ddr3}{DDR3}{Double Data Rate}
\newacronym{dff}{DFF}{Data Flip-Flop}
\newacronym{dll}{DLL}{Delay Locked Loop}
\newacronym{dmr}{DMR}{Dual Modular Redundancy}
\newacronym{dram}{DRAM}{Dynamic Random Access Memory}
\newacronym{dsd}{DSD}{Digital Synchronous Detection}
\newacronym{dsp}{DSP}{Digital Signal Processor}
\newacronym{dt}{DigiTime}{}
\newacronym{dvfs}{DVFS}{Dynamic Voltage and Frequency Scaling}
\newacronym{eda}{EDA}{Electronic Design Automation}
\newacronym{fdc}{FDC}{Frequency-to-Digital Converter}
\newacronym{fifo}{FIFO}{First In First Out}
\newacronym{fpga}{FPGA}{Field Programmable Gate Array}
\newacronym{gds}{GDS}{Global Data Share}
\newacronym{gnulgpl}{GNU LGPL}{GNU Lesser General Public Licence} 
\newacronym{gpgpu}{GPGPU}{General Purpose Graphics Processing Unit}
\newacronym{gpr}{GPR}{General Purpose Register}
\newacronym{gpu}{GPU}{Graphics Processing Unit}
\newacronym{gro}{GRO}{Gated Ring Oscillator}
\newacronym{io}{IO}{Input-Output}
\newacronym{hamsoc}{HAMSoC}{Hierarchical Agent Monitoring System-on-Chip}
\newacronym{hdl}{HDL}{Hardware Description Language}
\newacronym{hmp}{HMP}{Heterogeneous Multi-Processor}
\newacronym{ic}{IC}{Integrated Circuit}
\newacronym{icap}{ICAP}{Internal Configuration Access Port}
\newacronym[longplural={Intellectual Properties}]{ip}{IP}{Intellectual Property}
\newacronym{isa}{ISA}{Instruction Set Architecture}
\newacronym{lds}{LDS}{Local Data Share}
\newacronym{lru}{LRU}{Least Recently Used}
\newacronym{lsb}{LSB}{Least-Significant Bit}
\newacronym{lsu}{LSU}{Load Store Unit}
\newacronym{lut}{LUT}{Look Up Table}
\newacronym{mash}{MASH}{Multi-Stage Noise-Shaping}
\newacronym{mems}{MEMS}{Micro-Electro-Mechanical Systems}
\newacronym{miaow}{MIAOW}{Many-core Integrated Accelerator Of deepwater/Wisconsin}
\newacronym{mosfet}{MOSFET}{Metal Oxide Semiconductor Field Effect Transistor}
\newacronym{mpsoc}{MPSoC}{Multi-Processor System-on-Chip}
\newacronym{mshr}{MSHR}{Miss Status Holding/Handling Register}
\newacronym{noc}{NoC}{Network-on-Chip}
\newacronym{opencl}{OpenCL}{Open Computing Language}
\newacronym{ocn}{OCN}{On-Chip Network}
\newacronym{pcb}{PCB}{Printed Circuit Board}
\newacronym{pcie}{PCIe}{Peripheral Component Interconnect Express}
\newacronym{pl}{PL}{Programmable Logic}
\newacronym{pli}{PLI}{Verilog Programming Language Interface}
\newacronym{pll}{PLL}{Phase-Locked Loop}
\newacronym{ps}{PS}{Processing System}
\newacronym{pv}{PV}{Process Variation}
\newacronym{qoe}{QoE}{Quality of Experience}
\newacronym{qos}{QoS}{Quality of Service}
\newacronym{ram}{RAM}{Random Access Memory} 
\newacronym{risc}{RISC}{Reduced Instruction Set Computer}
\newacronym{riscv}{RISC-V}{Reduced Instruction Set Computing - V}
\newacronym{rtl}{RTL}{Register-Transfer Level}
\newacronym{sdk}{SDK}{Software Development Kit}
\newacronym{seec}{SEEC}{SElf-awarE Computing}
\newacronym{sgpr}{SGPR}{Scalar General Purpose Register}
\newacronym{si}{SI}{Southern Island}
\newacronym{simd}{SIMD}{Single Instruction Multiple Data}
\newacronym{simf}{SIMF}{Single Instruction Multiple Floating point}
\newacronym{sm}{SM}{Streaming Multiprocessor}
\newacronym{snr}{SNR}{Signal to Noise Ratio}
\newacronym[plural=SoCs,firstplural=Systems on Chip (SoCs)]{soc}{SoC}{System-on-Chip}
\newacronym{spared}{SPARED}{Self-aware PArtial Reconfiguration architecture for Edge Devices}
\newacronym{spice}{SPICE}{Simulation Program With Integrated Circuit Emphasis}
\newacronym{tad}{TAD}{Time \gls{adc}}
\newacronym[plural=TDCs,longplural={Time-to-Digital Converters}]{tdc}{TDC}{Time-to-Digital Converter}
\newacronym{tq}{TQ}{Time-Quantizer}
\newacronym{uart}{UART}{Universal Asynchronous Receiver/Transmitter}
\newacronym{vcdu}{VCDU}{Voltage Controlled Delay Unit}
\newacronym{vco}{VCO}{Voltage Controlled Oscillator}
\newacronym{vga}{VGA}{Video Graphics Array}
\newacronym{vhdl}{VHDL}{Very High Speed Integrated Circuit Hardware Description Language}
\newacronym{vlsi}{VLSI}{Very Large Scale Integration}
\newacronym{vgpr}{VGPR}{Vector General Purpose Register}
\newacronym{xilffs}{XILFFS}{Generic Fat File System Library}

\newacronym{amd}{AMD}{Advanced Micro Devices}
\newacronym{beol}{BEOL}{Back End Of Line}
\newacronym{cad}{CAD}{Computer-Aided Design}
\newacronym{cas}{CAS}{Circuits and Systems}
\newacronym{eu}{EU}{European Union}
\newacronym{fdd}{FDD}{fault detection and diagnostic}
\newacronym{fefet}{FeFET}{Ferroelectric Field Effect Transistor}
\newacronym{feline}{FeLINe}{FeFET Logic IN mEmory}
\newacronym[plural=FOMs,longplural={Figures of Merit}]{fom}{FoM}{Figure of Merit}
\newacronym{hipeac}{HiPEAC}{High Performance and Embedded Architecture and Compilation}
\newacronym{hp}{HP}{Hewlett Packard}
\newacronym{hqp}{HQP}{Highly Qualified People}
\newacronym{hvac}{HVAC}{Heating, Ventilation and Air Conditioning}
\newacronym{ibm}{IBM}{International Business Machines corporation}
\newacronym{ict}{ICT}{Institute for Computer Technology}
\newacronym{iot}{IoT}{Internet of Things}
\newacronym{kcl}{KCL}{Kirchhoff's circuit laws}
\newacronym{nda}{NDA}{Non-Disclosure Agreement}
\newacronym{nvp}{NVP}{Non-Volatile Processor}
\newacronym{oecd}{OECD}{Organization for Economic Cooperation and Development}
\newacronym{rd}{R\&D}{Research and Development}
\newacronym{soa}{SoA}{State-of-the-Art}
\newacronym{tsmc}{TSMC}{Taiwan Semiconductor Manufacturing Company}
\newacronym{tvlsi}{TVLSI}{Transactions on Very Large Scale Integration}

\def\diff{\mathrm{d}}

\def\xbarsize{$128\times 128$}
\def\Vset{\ensuremath{V_\mathrm{set}}}
\def\Vcond{\ensuremath{V_\mathrm{cond}}}
\def\Vclear{\ensuremath{V_\mathrm{reset}}}
\def\Vread{\ensuremath{V_\mathrm{read}}}
\def\von{\ensuremath{v_\mathrm{on}}}
\def\voff{\ensuremath{v_\mathrm{off}}}
\def\kon{\ensuremath{k_\mathrm{on}}}
\def\koff{\ensuremath{k_\mathrm{off}}}
\def\alphon{\ensuremath{\alpha_\mathrm{on}}}
\def\alphoff{\ensuremath{\alpha_\mathrm{off}}}
\def\Ron{\ensuremath{R_\mathrm{on}}}
\def\Roff{\ensuremath{R_\mathrm{off}}}
\def\won{\ensuremath{w_\mathrm{on}}}
\def\woff{\ensuremath{w_\mathrm{off}}}
\def\aon{\ensuremath{a_\mathrm{on}}}
\def\aoff{\ensuremath{a_\mathrm{off}}}
\def\RG{\ensuremath{R_\mathrm{G}}}
\let\Rg\RG
\def\fon{\ensuremath{f_\mathrm{on}}}
\def\foff{\ensuremath{f_\mathrm{off}}}

\def\konq{\ensuremath{k_{\mathrm{on},Q}}}
\def\koffq{\ensuremath{k_{\mathrm{off},Q}}}
\def\konp{\ensuremath{k_{\mathrm{on},P}}}
\def\koffp{\ensuremath{k_{\mathrm{off},P}}}
\def\vonq{\ensuremath{v_{\mathrm{on},Q}}}
\def\voffq{\ensuremath{v_{\mathrm{off},Q}}}
\def\vonp{\ensuremath{v_{\mathrm{on},P}}}
\def\voffp{\ensuremath{v_{\mathrm{off},P}}}
\def\Roffp{\ensuremath{R_{\mathrm{off},P}}}
\def\Ronp{\ensuremath{R_{\mathrm{on},P}}}
\def\Roffq{\ensuremath{R_{\mathrm{off},Q}}}
\def\Ronq{\ensuremath{R_{\mathrm{on},Q}}}
\def\Vq{\ensuremath{V_Q}}
\def\Vp{\ensuremath{V_P}}
\def\Vqi{V_{Q\mathrm{i}}}
\def\Vpf{V_{P\mathrm{f}}}
\def\Rq{\ensuremath{R_Q}}
\def\Rp{\ensuremath{R_P}}
\def\denom{\Rp\RG+\Rp\Rq+\Rq\RG}
\def\Rtext#1{\ensuremath{R_\mathrm{#1}}}
\def\RIH{\Rtext{IH}}
\def\RIL{\Rtext{IL}}
\def\ROH{\Rtext{OH}}
\def\ROL{\Rtext{OL}}

\begin{document}
	\def\mytitle{Device Variability Analysis for \\	{Memristive} Material Implication}
	\title{\mytitle}
	\author{Simon~Michael~Laube and Nima~TaheriNejad%
		\thanks{S.~M.~Laube and N.~TaheriNejad are with the TU~Wien, 1040 Vienna, Austria}
		\thanks{This work has been submitted to the IEEE for possible publication. Copyright may be transferred without notice, after which this version may no longer be accessible.}
		\thanks{}}
	\markboth{}{N.~TaheriNejad{} and S.~M.~Laube: \mytitle}
	\maketitle
	
	\begin{abstract}
		Currently, memristor devices suffer from variability between devices and from cycle to cycle. In this work, we study the impact of device variations on memristive \gls{imply}. New constraints for different parameters and variables are analytically derived and compared to extensive simulation results, covering single gate and 1T1R crossbar structures. We show that a static analysis based on switching conditions is not sufficient for an overall assessment of robustness against device variability. Furthermore, we outline parameter ranges within which the \gls{imply} gate is predicted to produce correct output values. Our study shows that threshold voltage is the most critical parameter.	This work helps scientists and engineers to understand the pitfalls of designing reliable \gls{imply}-based calculation units better and design them with more ease. Moreover, these analyses can be used to determine whether a certain memristor technology is suitable for implementation of \gls{imply}-based circuits and systems. 
	\end{abstract}
	\section{Introduction}\label{sec:intro}
Memristors are used for memory applications~\cite{Niu2010, Ho2011, Mohammad2013, Baghel2015, Radakovits2019}, where even storage of multiple bits per device is feasible~\cite{zangeneh_design_of_1t1r_reram,kim2010cnna,Taherinejad2015ems,Taherinejad2016cce}. In addition, memristors have become increasingly popular for neural network and learning applications~\cite{Pershin2012ieee,Thomas2013applied}, by exploiting their analog, synapses-like nature. Another application of memristors is implementing digital (in-memory) logic~\cite{borghetti_memristive_stateful_logic, talati_logic_design_magic, CRS_proposal}, such as \gls{imply}, for various computations~\cite{Gupta2018, Papandroulidakis2017, Rohani2017, Taherinejad2019newcas, guckert2018system, Radakovits2020tcasi}. At the moment these applications -- more often than not -- are 
not verified by physical implementation and experimental data~\cite{CAS_physRealization}. This imbalance leads to many problems when actual physical implementation is desired. While material sciences have certainly progressed in this field~\cite{waser_redox_reram_physics,menzel_switching_kinetics_ecm,menzel_reram_physics}, the circuit-level interface to higher abstraction levels is not yet ready to provide a reliable base for proposed applications~\cite{CAS_physRealization}. Some of the fundamental problems, that need to be considered at design time, are {inter-device variability}  and {cyclic variability}. In larger structures, usually implemented within crossbar arrays, sneak paths and wire resistance are an even bigger issue~\cite{cassuto_sneak_path_constraints,sneak_paths_closed_form}. While the two latter have received an acceptable level of attention from the community, the two former have been less explored and addressed by the community. We hope that this work encourages and provides pointers to the community to move in that direction.\par 

Here, we aim to provide a better insight into the operation of a single memristive \gls{imply} logic gate by considering device variations. Similar works on that topic already exist, such as~\cite{kvatinsky_device_variations,xie_robustness_of_memristor_logic,chen_imply_ron_not_reached}, which mainly focus on other types of memristor-based logic. The most relevant work to ours is~\cite{chen_imply_ron_not_reached}, where the focus is set on the design of the \gls{imply} circuit itself, and an alternative operation  ``\gls{nimp}'' is proposed to mitigate certain problems. In contrast, our work explores regular \gls{imply} in more detail, particularly regarding the effect of device variations on the \gls{imply} operation, and leaves the \gls{nimp} approach for future works. We note that there is a variety of memristors based on different physical effects~\cite{waser_redox_reram_physics}. From the perspective of this work, the internal mechanism is to a large extent inconsequential. Hence, we use the general term, memristor, to refer to \emph{resistive switching elements} or \glspl{reram} and, when needed, specify what may make the internal mechanisms important. 
The main contribution of this work to the field of memristor-based logic, is an in-depth mathematical analysis of memristive \gls{imply} regarding its constraints due to device variation. Plausibility of the proposed constraints is verified via simulations using a popular model. \par 

The rest of this paper is organized as follows: \Cref{sec:imply_logic}, particularly \Cref{subsec:gate_and_constr}, reviews memristive \gls{imply} logic and shows its limitations. An introduction to the crossbar architecture is given in \Cref{sec:xbar_fundamental} and the device model is described in \Cref{sec:model}. 
In \Cref{sec:math}, we formulate new constraints for the \gls{imply} gate, before comparing them to the single gate simulation results in \Cref{sec:sim_single}. The results of the crossbar simulations are presented in \Cref{sec:sim_xbar} and compared against the single gate simulation and constraints. We conclude the paper in \Cref{sec:conclusion}. 

\section{Material implication (IMPLY)}\label{sec:imply_logic}
The truth table of \gls{imply}, and its four different cases, are shown in~\Cref{tab:truth_table_imply}. It takes two input states $p$ and $q$ and outputs $q'$. Not every type of memristor is suitable for material implication. The devices have to exhibit \emph{voltage threshold behavior}. Moreover, all devices used for an operation shall have the same parameter values (resistance range, threshold voltages, switching speed).

\begin{table}
	\centering
	\caption{Truth table of material implication and its four cases}
	\label{tab:truth_table_imply}
	\begin{tabular}{c|cc|c}
		\hline\hline
		Cases & $p$ & $q$ & $q'$\\
		\hline
		{Case} 1 &0 & 0 & 1 \\
		{Case} 2 &0 & 1 & 1 \\
		{Case} 3 &1 & 0 & 0 \\
		{Case} 4 &1 & 1 & 1 \\
		\hline\hline
	\end{tabular}
\end{table}

\subsection{Gate structure and constraints}\label{subsec:gate_and_constr}
\begin{figure}[!b]
	\centering
	\subfigure[]{
		\begin{circuitikz}
			\draw(0,0) to[memristor,i=$i$] (2,0);
			\draw[-stealth](.2,.5)node[above]{$+$} -- (1,.5)node[above]{$v$} -- (1.8,.5)node[above]{$-$};
			\path(0,0) -- (0,-1.5);
		\end{circuitikz}
		\label{fig:symbol_polarity}
	}
	\subfigure[]{
		\tikzset{varr/.style={-stealth}}
		\begin{circuitikz}[scale=.6,transform shape]
			\draw (.7,-1.5) to[short] (.7,-1)
			(3.3,-1.5) to[short] (3.3,-1) 
			(3.3,-3.5) -- (-1,-3.5) to[R,l_=$R_G$,font=\Large] (-1,-5) node[ground]{};
			\draw(.7,-3.5) to[Mr,*-o,l^=$P$,font=\Large] (.7,-1.5) node[left]{$R$}
			(3.3,-3.5) to[Mr,-o,l^=$Q$,font=\Large](3.3,-1.5) node[left]{$T$};
			\draw(-.5,-1) rectangle (4.5,0)
			(2,-.5) node[font=\Large] {Driver \& Sensing};
			\draw[varr] (1.2,-1.75)node[right]{$+$} --(1.2,-2.5) node[right]{\Large$V_P$}-- (1.2,-3.25)node[right]{$-$};
			\draw[varr] (3.8,-1.75)node[right]{$+$} --(3.8,-2.5) node[right]{\Large$V_Q$}-- (3.8,-3.25)node[right]{$-$};
			\draw[varr] (-.5,-3.7)node[right]{$+$} --(-.5,-4.5) node[right]{\Large$V_G$}-- (-.5,-5.3)node[right]{$-$};	
		\end{circuitikz}
		\label{fig:imply_gate}}	
	\caption{\subref{fig:symbol_polarity} Memristor symbol and defined voltage polarity used in this work. \subref{fig:imply_gate} A single IMPLY gate.}
	\label{fig:symbol_and_imply_gate}
\end{figure}
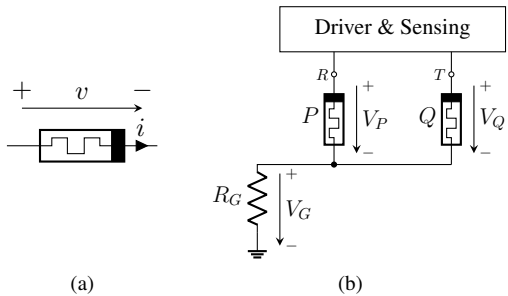

Two memristors and a resistor are necessary for a single memristive \gls{imply} gate. \Cref{fig:imply_gate} shows such a gate, with abstracted drivers and sense circuitry. Each memristor can be \textit{set} (forced to \gls{lrs}) by applying a voltage $|\Vset|>|\von|$ with appropriate (in our case \textit{negative}) polarity; and can be \textit{reset} (forced to \gls{hrs}) by applying $|\Vclear|>|\voff|$ with an opposite polarity\footnote{This is true for bipolar switching mechanisms. \gls{pc} based devices, for example, may use the same voltage polarity for set and reset.}. If a memristor is set, it represents logic state `1'; if it is reset, it represents logic state `0'~\cite{borghetti_memristive_stateful_logic}. During initialization, these voltage amplitudes are applied to each device, while the other is kept floating. For the actual logic operation both devices are driven at the same time: $\Vcond$ is applied to node $R$ and $\Vset$ to node $T$ of \Cref{fig:imply_gate}. 
For a correct operation 
\begin{align}
|\Vset| > |\von| \label{eqn:vset_relation}\\
|\Vset - \Vcond| < |\von| \label{eqn:vcond_relation1}\\
|\Vclear| > |\voff| \label{eqn:vclear_relation}
\end{align}
must hold. Moreover, the circuit designer needs to select a valid value for $\RG$, as described in~\cite{kvatinsky_imply_logic_design}.\par 

 It is important to note that only in Case~1 of \Cref{tab:truth_table_imply} the output memristor $Q$ is actually changing its state. However, during this process the voltage across each device changes too. It is valid to ask if this has an effect on the result of the operation, and the answer is yes. Using \gls{kcl}, Chen et al.~\cite{chen_imply_ron_not_reached} showed that there are two possible final steady states of the operation:
 \begin{enumerate}
 	\item The normalized state variable $s$ reaches the upper boundary of $1$ ($R_Q=\Ron$) before the voltage across $Q$ falls below the threshold $\von$. The final steady state is $R_Q=\Ron$.
 	\item $\Vq$ falls below the threshold $\von$ before $s$ reaches $1$. In this case the steady state resistance can be expressed as\footnote{Note that in our convention $\von<0$.}:
 	\begin{align}
 	R_\mathrm{min} &= \frac{-\von\,\RG\,\Roff}{(\RG+\Roff) (\Vset+\von) - \RG \Vcond}\label{eqn:Rmin}
 	\end{align}
 \end{enumerate}
 An important point to mention is that this calculation is based on the premise, that the driving voltages of $P$ and $Q$ are chosen such that there is \emph{no state drift in $P$} during operation. 
	\subsection{Crossbar principles}\label{sec:xbar_fundamental} 
Crossbar architectures are a natural candidate for memristor-based logic, as high integration density can be reached. 
In so called 1R (or 1M) crossbars, a memristor device is fabricated at each intersection of bit- and word-lines, which act as the access medium for the cell. 1R crossbars are very difficult to handle~\cite{wan2010edl,li2018imw,li2018nature}, even if parasitics are not considered. Many works have been carried out to study effects~\cite{cassuto_sneak_path_constraints,chen_crossbar_array_model,shin_data_dependent_statistical_model_analysis,shin_data_dependend_model}, or solve them~\cite{chen_crossbar_array_model,sneak_paths_closed_form, CRS_proposal}, but thus far 1T1R has been the preferred implementation~\cite{wan2010edl,li2018imw,li2018nature}. 1T1R (or 1T1M) crossbars,consist of a transistor and a memristor in each cell. The transistor in each cell cost extra area but they prevent the cells from switching state when a cell is not part of an operation (not selected). One possible, readout scheme is provided by~\cite{sneak_paths_closed_form}, which we use in this work. The chosen readout scheme~\cite{sneak_paths_closed_form} provides a closed-form solution. 
Moreover, it introduces very little additional complexity, which enables this work to remain focused on issues regarding \gls{imply} itself. In \Cref{sec:sim_xbar} we compare crossbar simulation results against single gate results and outline the differences.

\section{Device model}\label{sec:model}
There are a range of different simulation models for memristors~\cite{kvatinsky_team,kvatinsky_vteam,strachan_memristor_model,jiang_stanford_model}. For the simulations presented in this paper, the TU Wien LTSpice implementation~\cite{VTEAM_Spice, CAS_physRealization} of VTEAM~\cite{kvatinsky_vteam} was used. An overview of the model is given in \Cref{eqn:w,eqn:s,eqn:dwdt,eqn:tu_ui,eqn:s_ui}, with a memristor polarity as shown in \Cref{fig:symbol_polarity}. \par 

In VTEAM, $w$ acts as the state variable and represents a length between the extrema $\won$ and $\woff$ ($w\in[\woff,\won]$). 
Here, we define the relation of these state variable boundaries
\begin{align}
\won &> \woff\label{eqn:w}
\end{align}
and define the normalized state variable ($s\in[0,1]$): 
 \begin{align}
s(w)= w' &= \frac{w-\woff}{\won-\woff}\label{eqn:s}
\end{align}	
These definitions may be changed, as long as the model equations are updated, too.
The rate of change of the state variable, $w$, is defined by 
\begin{align}
\frac{\diff w}{\diff t} &= \begin{cases}
\koff \left(\frac{v}{\voff}-1\right)^{\alphoff} \foff(w)\ \ &0<\voff<v\\
0 & \von < v < \voff\\
\kon \left(\frac{v}{\von}-1\right)^{\alphon} \fon(w) & v<\von<0\\
\end{cases}\label{eqn:dwdt}
\end{align}	
which is the essential building block of the model~\cite{kvatinsky_vteam}. In this equation $\kon,~\koff,~\von,~\voff,~\alphon$ and $\alphoff$ represent fitting parameters, while $\fon(w)$ as well as $\foff(w)$ are window functions that limit $\diff w/\diff t$. \par 

$I/V$-characteristics and window functions are not defined in the model and thus can be freely chosen. We chose a linear current/voltage dependency:
\begin{align}
R(w)=\Roff+\left(\Ron-\Roff\right)\cdot s(w)
\label{eqn:tu_ui}
\end{align}	
By rearranging the equation we can further express $s(R)$ for any (measured) $R$:
\begin{align}
s(R) &= \frac{R-\Roff}{\Ron - \Roff}\label{eqn:s_ui}
\end{align}

The same expressions as for the Simmon's Tunnel Barrier model in~\cite{kvatinsky_team} were chosen as window functions. In addition, $w$ is bounded and thus cannot exceed $\won$ or $\woff$. \par 

The studies presented in this paper are kept as general as possible, however, simulations need model parameters. Rather than introducing arbitrary parameter values, we experimentally fitted~\cite{Taherinejad2019newcas} our VTEAM model to Knowm BS-AF-W~\cite{knowm_datasheet} memristors we had at the time.
Parameters shown in \Cref{tab:parameters}, represent a best effort fitting we conducted previously~\cite{Taherinejad2019newcas}.

\section{Formulating Constraints}\label{sec:math}
This section marks the beginning of our new contributions.	In this section, we mathematically extract device variability constraints which govern and limit operations of \gls{imply}. At first, we define the notation: Each parameter involved in the analysis is written as $\xi_{i,M}$, where $\xi\in\{R,v,k\}$, $M \in \{P, Q\}$ and $i \in \{\mathrm{off}, \mathrm{on}\}$.
For example, the off-resistance of memristor $P$ in this notation would be denoted as $\Roffp$. \par 

Logic thresholds determine the logic state of a device. They are defined separately for input (I) and output (O), as well as logic `1' (H) and `0' (L). Indices are used to denote the respective logic thresholds, e.g. $\RIL$ is the input threshold for logic `0'.

\subsection{Static behavior}\label{subsec:stat_math}
Each case in the truth table (\Cref{tab:truth_table_imply}) imposes constraints onto the voltage $\Vq$ across memristor $Q$, as certain \textit{switching conditions} must be met. They can be analyzed via \gls{kcl} and  represent a \textit{static} view of the circuit. The constraints can be used to find limits for $\Ronp$, $\Roffp$, $\vonq$ and $\voffq$. They do not provide limits for $\Ronq$ or $\Roffq$, as $R_Q$ in this context is the target output resistance state that $Q$ must reach during \gls{imply}. Therefore, $R_Q$ is later set according to the chosen output logic threshold: $R_Q\leq R_\mathrm{OH}$ or $R_Q\geq R_\mathrm{OL}$.Applying \gls{kcl} in \Cref{fig:imply_gate} gives us the voltage across $Q$ as
\begin{align}
	\Vq &= \frac{\Rq (\Rp+\RG)\Vset-\Rq\RG\Vcond}{\denom}\label{eqn:VQ}.
\end{align}
First we solve \Cref{eqn:VQ} with a generalized threshold voltage, $v$, and the solution is specialized for each case afterwards.
The first switching condition is:
\begin{align}
\Vq&> v.
\label{eqn:VQ_constr}
\end{align}
Plugging \Cref{eqn:VQ} into \Cref{eqn:VQ_constr} and isolating $\Rp$ leads to 
\begin{align}
\Rp\cdot b&> a\label{eqn:Rp_intermediate},
\end{align}
where
\begin{align}
a= \Rq\RG (v&+\Vcond-\Vset)\label{eqn:a},\\
b=	\Rq\Vset-&v(\RG+\Rq)\label{eqn:b}.
\end{align}
At this point the relation in \ref{eqn:Rp_intermediate}, is divided by $b$. Therefore, depending on the value of $b$, we have
\begin{align}
\begin{cases}
		\Rp>\frac{a}{b}\ &\text{if}\ \ \ b > 0\\
		\Rp<\frac{a}{b}\ &\text{if}\ \ \ b < 0\\
		\Rp\rightarrow\pm\infty\ &\text{if}\ \ \ b=0
\end{cases}\label{eqn:Rp_case}
\end{align}
Next, the switching condition
\begin{align}
   \Vq&< v
\end{align}
is examined. Following the same steps as before, we have 
\begin{align}
\begin{cases}
	\Rp<\frac{a}{b}\ &\text{if}\ \ \ b > 0\\
	\Rp>\frac{a}{b}\ &\text{if}\ \ \ b < 0\\
	\Rp\rightarrow\pm\infty\ &\text{if}\ \ \ b=0
\end{cases}\label{eqn:Rp_case2}
\end{align}
Since the third case ($b=0$) in \Cref{eqn:Rp_case,eqn:Rp_case2} 
yields\footnote{That is, as long as $|a|$ is neither zero, nor $\infty$.} $\pm\infty$, it is of no interest for the rest of the analysis.
The first two cases in \Cref{eqn:Rp_case,eqn:Rp_case2} both provide limits for $v$ and $\Rp$, respectively. \par 

Here, the resulting equations (constraints) are specialized for each of the four cases of the truth table using the respective switching conditions. {$\Rq$ is set to the associated output logic threshold ($\ROH$ or $\ROL$).} Only the first case ($b>0$) of \Cref{eqn:Rp_case,eqn:Rp_case2} is considered, since the second case ($b<0$) only provides negative limits, and $\Rp >0$.  
For every case of the truth table, according to our notations, $\von<0$ and $\voff>0$. Hence, we have
\begin{description}
	\item[Case 1] 	$\Vq > -\vonq$
	\begin{align}
	\vonq&> -\Vset\frac{\ROH}{\RG+\ROH}\label{eqn:vonq_case1}\\
	\Roffp &> \frac{\ROH\RG (\Vcond-\vonq-\Vset)}{\ROH\Vset+\vonq (\RG+\ROH)}\label{eqn:Roffp_case1}
	\end{align}
	\item[Case 3:] 	$\Vq < -\vonq$
	\begin{align}
	\vonq&> -\Vset\frac{\ROL}{\RG+\ROL}\label{eqn:vonq_case3}\\
	\Ronp &< \frac{\ROL\RG (\Vcond-\vonq-\Vset)}{\ROL\Vset+\vonq (\RG+\ROL)}\label{eqn:Ronp_case3}
	\end{align}
	
	\item[Case 2/Case 4:] $\Vq > -\voffq$
	\begin{align}
	\voffq&> -\Vset\frac{\ROH}{\RG+\ROH}\label{eqn:voffq_case24}\\
	\Roffp &> \frac{\ROH\RG (\Vcond-\voffq-\Vset)}{\ROH\Vset+\voffq (\RG+\ROH)}\label{eqn:Roffp_case2}\\
	\Ronp &> \frac{\ROH\RG (\Vcond-\voffq-\Vset)}{\ROH\Vset+\voffq (\RG+\ROH)}\label{eqn:Ronp_case4}
	\end{align}
\end{description}
\Cref{eqn:vonq_case1,eqn:vonq_case3,eqn:voffq_case24} directly result from $b>0$, whereas \Cref{eqn:Roffp_case1,eqn:Ronp_case3,eqn:Roffp_case2,eqn:Ronp_case4} are the respective relations derived from $R_P>a/b$ in \Cref{eqn:Rp_case} and $R_P<a/b$ in \Cref{eqn:Rp_case2}.\par 

Some additional static constraints are given by the choice of logic thresholds. That is, 
\begin{align}
\Roffp > R_\mathrm{IL}\label{eqn:Roffp_greater_RIL}\\
\Roffq > R_\mathrm{IL}\label{eqn:Roffq_greater_RIL}\\
\Ronp < R_\mathrm{IH}\label{eqn:Ronp_smaller_RIH}\\
\Ronq < R_\mathrm{IH}\label{eqn:Ronq_smaller_RIH}.
\end{align}
Similar to standard logic families, input and output thresholds may differ. From the point-of-view of these constraints, only input thresholds need to be considered, as they determine whether or not the device states fed to the operation are valid in the first place.

\subsection{Dynamic behavior}\label{subsec:dyn_math}
With respect to the static analysis, the chosen timestep of operation can introduce much stricter constraints. For exact solutions, one would have to solve the differential state equation of the chosen model (in our case \Cref{eqn:dwdt} from VTEAM). This is not a trivial task and might not even be possible for all models. Thus, in this section we derive a lower boundary for $\vonq$, but not the infimum, which cannot be exceeded by the exact (or numeric) solution. That way we take into account the state change (dynamic behavior) of memristors during the operation, using an acceptable estimation. We note that in doing such an analysis, the chosen model is assumed to be accurate. However, in practice no existing model represents all the reality and physics involved. \par 

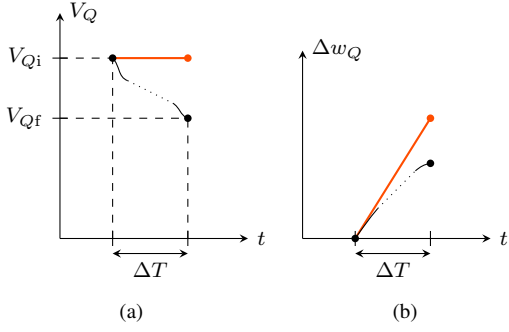
\begin{figure}
	\centering
	\subfigure[]{
		\begin{tikzpicture}[font=\footnotesize]
		\draw[-stealth] (0,0) -- (2.5,0) node[right]{$t$};
		\draw[-stealth] (0,0) -- (0,3) node[right]{$\Vq$};
		\draw[] (.7,.1) --(.7,-.1) 
		(1.7,.1) --(1.7,-.1)
		(.1,2.4) -- (-.1,2.4) node[left]{$\Vqi$}
		(.1,1.6) -- (-.1,1.6) node[left]{$V_{Q\mathrm{f}}$};
		\draw[stealth-stealth] (.7,-.2) -- (1.2,-.2) node[below] {$\Delta T$} -- (1.7,-.2);
		\draw[thick, red!70!yellow] (.7,2.4) -- (1.7,2.4);
		\fill[red!70!yellow] (1.7,2.4)circle(.05);
		\draw[dashed] (.7,0) -- (.7,2.4)
		(1.7,0) --(1.7,1.6)
		(0,2.4) -- (.7,2.4)
		(0,1.6)--(1.7,1.6);
		\fill[black] (.7,2.4) circle(.05)
		(1.7,1.6) circle(.05);
		\draw (.7,2.4) to[out=-45, in=160] (.9,2.1)
		(1.7,1.6) to[out=160, in=-30] (1.5,1.8);
		\draw[dotted] (.9,2.1)--(1.5,1.8);
		\end{tikzpicture}
		\label{fig:vq_during_imply}}
	\subfigure[]{	
		\begin{tikzpicture}[font=\footnotesize]
		\draw[-stealth] (0,0) -- (2.5,0) node[right]{$t$};
		\draw[-stealth] (0,0) -- (0,2.5) node[right]{$\Delta w_Q$};
		\draw[] (.7,.1) --(.7,-.1) 
		(1.7,.1) --(1.7,-.1);
		\draw[stealth-stealth] (.7,-.2) -- (1.2,-.2) node[below] {$\Delta T$} -- (1.7,-.2);
		\draw[thick, red!70!yellow] (.7,0) -- (1.7,1.6);
		\fill[red!70!yellow] (1.7,1.6)circle(.05);
		\fill[black] (.7,0) circle(.05)
		(1.7,1) circle(.05);
		\draw (.7,0) to[out=55, in=-130] (1,.4)
		(1.7,1) to[out=-170, in=40] (1.5,.9);
		\draw[dotted] (1,.4)--(1.5,.9);
		\end{tikzpicture}
		\label{fig:dw_during_imply}}
	\caption{A symbolic voltage-time curve for $\Vq$ \subref{fig:vq_during_imply} an induced state change $\Delta w_Q$ \subref{fig:dw_during_imply} during a single \gls{imply} operation of duration $\Delta T$. The estimations for the formulation of constraints are drawn in orange, the symbolic representations of the actual curves in black.}
	\label{fig:changes_during_imply}
\end{figure}
The main idea of our estimation is to look at how $\Vq$ changes over time in Case 1 of the truth table, while assuming negligible state drift in $P$. As $\Rq$ changes from  \gls{hrs} to \gls{lrs}, $\Vq$ decreases. Thus, the initial voltage $\Vqi$ is the highest occurring value of $\Vq$ during that timestep, while the final voltage $V_{Q\mathrm{f}}$ is the lowest -- symbolically shown in \Cref{fig:vq_during_imply}. If the device characteristics are such that the highest $\Vq$ corresponds to the maximum value of $\diff w/ \diff t$ -- in our case true due to \Cref{eqn:dwdt} -- a hard limit can be expressed. \gls{kcl} can be used to describe the initial voltage
\begin{align}
\Vqi &= \frac{\Roffq(\Roffp+\Rg)\Vset-\Roffq \Rg \Vcond}{\Roffp \Rg+ \Roffp \Roffq + \Roffq \Rg}.
\label{eqn:VQi}
\end{align}
 Plugging $\Vqi$ into \Cref{eqn:dwdt} gives the inital rate of state change\footnote{$\fon$ is missing in \Cref{eqn:dwqdt_initial} because 
 $\fon\approx 1$ for $w_Q < \won$}:
\begin{align}
\frac{\diff w_Q}{\diff t}\bigg|_{\text{initial}} &= \frac{\Delta w_Q}{\Delta T}= \konq \left(\frac{-\Vqi}{\vonq}-1\right)^\alpha\label{eqn:dwqdt_initial}
\end{align} 
Now we set the actual $\diff w_Q/\diff t$ equal to the initial rate for the whole timestep $\Delta T$. Through this simplification a maximum $\Delta w_Q$ for the given timestep $\Delta T$ can be found, which cannot be exceeded:
\begin{align}
\Delta w_Q &=  \konq \left(\frac{-\Vqi}{\vonq}-1\right)^\alpha \Delta T\label{eqn:Dwq_max}
\end{align}
 This is because the estimation provides a better overall situation towards the correct operation result, when compared to the actual situation. That is, as we see in \Cref{fig:dw_during_imply}, the estimated $\Delta w_Q$ is always larger than the actual value.
To obtain a correct result after the IMPLY operation, $\Rq$ must at least reach the logic threshold $R_\mathrm{OH}$. Otherwise the result would not be interpreted as logic `1'. Via \Cref{eqn:s_ui} we can find $s(R_\mathrm{OH})$. In combination with \Cref{eqn:s}, the necessary $\Delta w_\mathrm{min}$ can be expressed as
\begin{align}
\Delta w_\mathrm{min}&= \frac{R_\mathrm{OH}-\Roffq}{\Ronq-\Roffq}(\won-\woff) + \woff,
\label{eqn:wmin}
\end{align}
and 
\begin{align}
\Delta w_Q &\geq \Delta w_\mathrm{min}\label{eqn:Dwq_greater_Dwmin}
\end{align}
shall be true. Plugging the previous terms into \Cref{eqn:Dwq_greater_Dwmin} gives 
\begin{align}
 \vonq&\geq\frac{-\Vqi}{\sqrt[\leftroot{-1}\uproot{2}\scriptstyle\alpha]{\frac{\Delta w_\mathrm{min}}{\konq\Delta T}} +1}, \label{eqn:vonq_wmin}
\end{align}
which is the newly found constraint. 
As this relation contains multiple parameters of $P$ and $Q$ apart from $\vonq$, it provides boundaries for all of them. For example, a certain $\vonq$ restricts $\Roffp$ to a specific range, and in turn a certain $\Roffp$ restricts $\vonq$ to a specific range. \par 

Considering \Cref{eqn:vonq_wmin}, a question is whether the
same estimation could be used to find an upper limit for $\vonq$. Such an analysis, however, is not meaningful for memristor $Q$. Both, a minimum value of $\diff w/ \diff t$ and a maximum value $\Delta w_\mathrm{max}$, must be specified. The later does not exist for $Q$ since a high $w_Q$ (ideally $\won$) is desired in Case~1.\par 

There is, however, a $\Delta w_\mathrm{max}$ for memristor $P$, as a change of $\Rp$ is generally not desired. By definition the logic state of $P$ remains unchanged if $\Rp>R_\mathrm{IL}$. Only if this is true, it can be used as an input for an operation. As $\Rp$ drifts away from $\Roffp$  (ideal \gls{hrs}), $\Vp$ decreases. Thus, the minimum value of $\diff w/\diff t$ is the \textit{final value} at the end of the operation, in contrast to $\Vqi$ being the initial value. The final value $\Vpf$ cannot be expressed easily, as $R_{P\mathrm{f}}$ and $R_{Q\mathrm{f}}$ are unknown. Hence, another simplification must be made: We evaluate $\Vpf$ using $\Rp=\Roffp$, as if there  was no state drift in the first place:
\begin{align}
V_{P\mathrm{f},j} &= \frac{\Roffp(R_{Q,j}+\Rg)\Vcond-\Roffp\Rg\Vset}{\Roffp\Rg+\Roffp R_{Q,j}+R_{Q,j}\Rg}\label{eqn:vpf}
\end{align}
Due to this very rough estimation, we expect the constraint to represent a fairly weak boundary. Therefore, 
\begin{align}
R_{Q,1}&=R_{\mathrm{min},Q}\label{eqn:Rq1}\\
R_{Q,2}&=\frac{\Roffq+R_{\mathrm{min},Q}}{2}\label{eqn:Rq2}\\ 
R_{Q,3}&=\sqrt{\Roffq\cdot R_{\mathrm{min},Q}}\label{eqn:Rq3}
\end{align}
 are used to evaluate $\Vpf$ in \Cref{eqn:vpf} and obtain a range within which the circuit is less likely to fail. The first value, \Cref{eqn:Rq1}, is the theoretical minimum for $\Ronq$, which is the ideal $R_{Q\mathrm{f}}$. However, the actual $R_{Q,\mathrm{f}}$ can take on any value between $\Roffq$ and $R_{\mathrm{min},Q}$. Hence, in a second estimation, \Cref{eqn:Rq2}, we assume that $R_{Q\mathrm{f}}$ is the arithmetic mean of $\Roffq$ and $R_{\mathrm{min},Q}$. In other words, the final state is halfway between the initial state and the ideal endstate ($R_{\mathrm{min},Q}$). However, if $V_{P\mathrm{f}}$ is plotted over $R_Q$ on a linear scale, it reveals that the dependence is non-linear. Thus, $R_{Q,2}$ might not be the best estimation either. The dependence is, nevertheless, approximately linear on a semi-logarithmic scale; so our third estimation, \Cref{eqn:Rq3}, assumes $R_{Q\mathrm{f}}$ to be the geometric mean of $\Roffq$ and $R_{\mathrm{min},Q}$. If the timestep of \gls{imply} operation is limited, we do not expect $R_{Q,1}$ to provide an appropriate estimation, since this is the overall optimum scenario. $R_{Q,2}$ and $R_{Q,3}$ might both be of value to the circuit designer, because they represent a non-ideal scenario, chosen based on design parameters. \par 

Following the same steps as before, we can formulate the constraint for $\vonp$: 
\begin{align}
\Delta w_P &=  \konp \left(\frac{-V_{P\mathrm{f},j}}{\vonp}-1\right)^\alpha \Delta T\\
\Delta w_P &\leq \Delta w_\mathrm{max}\\
\vonp&\leq\frac{-V_{P\mathrm{f},j}}{\sqrt[\leftroot{-1}\uproot{2}\scriptstyle\alpha]{\frac{\Delta w_\mathrm{max}}{\konp\Delta T}} +1}\label{eqn:vonp_wmax}
\end{align}
\Cref{tab:constraints_summary} provides a summary of relevant constraints on memristor parameters, that were derived in this section. Most of these relations depend on multiple parameters of both memristors. Thus, the permissible value range of one parameter is impacted by the values of other parameters, and vice versa. Once the value of a parameter is determined (either decided by the designer or given by the technology) respective equations in \Cref{tab:constraints_summary} determine the tolerable range of variation in others. This bidirectional view enables us to define an operating area, which, in turn, allows us to predict how the circuit will react to variations in the concerned parameters. 

\begin{table}[t]
\scriptsize
\centering
\caption{Summary of related constraints on parameters of $P$ and $Q$.} 
\def\arraystretch{1.2}
\begin{tabular}{c|c}
\hline\hline
\multirow{2}{1cm}{Constraint} & Constrained \\
& parameters\\
\hline  
$\vonq>f(\Rq\equiv\ROH)$  & \multirow{2}{.6cm}{$\vonq$}  \\
\Cref{eqn:vonq_case1}&\\
\hline
$\Roffp>f(\vonq,\Rq\equiv\ROH)$ & \multirow{2}{1.4cm}{$\Roffp, \vonq$} \\
\Cref{eqn:Roffp_case1}&\\
\hline 
$\vonq>f(\Rq\equiv\ROL)$  & \multirow{2}{.6cm}{$\vonq$}  \\
\Cref{eqn:vonq_case3}&\\
\hline
 $\Ronp<f(\vonq,\Rq\equiv\ROL)$ & \multirow{2}{1.4cm}{$\Ronp, \voffq$} \\
\Cref{eqn:Ronp_case3}&\\
\hline
$\vonq>f(\Roffp,\Ronq,\Roffq,\konq)$  & $\Roffp,\Ronq,\Roffq,$  \\
 \Cref{eqn:vonq_wmin}& $\vonq,\konq$  \\
\hline
$\vonp<f(\Ronp,\Roffp,\Roffq,\konp)$ & $\Ronp,\Roffp,\Roffq,$  \\
\Cref{eqn:vonp_wmax}& $\vonp,\konp$ \\
\hline
$\Ronp<\RIH$, \Cref{eqn:Ronp_smaller_RIH} &$\Ronp$\\
$\Roffp>\RIL$, \Cref{eqn:Roffp_greater_RIL}& $\Roffp$ \\
$\Ronq<\RIH$, \Cref{eqn:Ronq_smaller_RIH} & $\Ronq$\\
$\Roffq>\RIL$, \Cref{eqn:Roffq_greater_RIL} & $\Roffq$\\
\hline\hline 
\end{tabular}
\label{tab:constraints_summary}
\end{table}

\section{Simulation -- Single Gate}\label{sec:sim_single}
\subsection{Circuit design}\label{subsec:sim_circuit}
The simulated circuit corresponds to the circuit shown in \Cref{fig:imply_gate}, with the addition that $\RG$ can be shorted by a parallel switch. The driver circuits are ideal voltage sources with serial switches for High-Z mode. Each switch is modeled with an on-resistance of $1\,\mathrm{n\Omega}$ and an off-resistance of $1\,\mathrm{G\Omega}$. Given the memristor properties, especially $\Ron$ and $\Roff$, five circuit-level parameters have to be determined. These are $\RG, \Vset, \Vcond, \Vclear$ and $\Vread$. Choosing $\Vclear$ is somewhat straightforward, as it is only used for initialization and not the \gls{imply} operation per s\'{e}. \par 

For this work, this voltage was set to $\Vclear=-1\,\mathrm{V}$.
Next, $\Vset$ and $\Vcond$ are determined. We define $\Vset=1\,\mathrm{V}, \Vcond=0.9\,\mathrm{V}$, based on the memristor's properties and~\Cref{eqn:vset_relation,eqn:vcond_relation1}. With the voltages set, the constraints on $\RG$~\cite{kvatinsky_imply_logic_design} can be evaluated, which leads to: $5.000\,\mathrm{k\Omega} < \RG < 230.769\,\mathrm{k\Omega}$. $\RG=40\,\mathrm{k\Omega}$ was chosen as the value of this resistor, which is lower than the geometric mean ($100\,\mathrm{k\Omega}$) proposed by~\cite{kvatinsky_imply_principles}. A summary of model and circuit parameters is shown in~\Cref{tab:parameters}, where the former are based on experimental results from previous works~\cite{Taherinejad2019newcas,semiparallel_imply_adder}. 

\begin{table}[t]
	\centering
	\caption{Nominal values of model and circuit parameters.}
	\label{tab:parameters}
	\scriptsize
	\begin{tabular}{l c c c c c c}
		\hline\hline
		Parameter &$\von$&$\voff$&$\Ron$&$\Roff$&\multicolumn{2}{c}{$\kon$}\\
		Value &$-0.7\,\mathrm{V}$&$10\,\mathrm{mV}$&$10\,\mathrm{k\Omega}$&$1\,\mathrm{M\Omega}$&\multicolumn{2}{c}{$1\,\mathrm{cm/s}$} \\
		\hline
		Parameter&$\alphon$&$\alphoff$&$\won$&$\woff$&\multicolumn{2}{c}{$\koff$}\\
		Value&3&3&$3\,\mathrm{nm}$&$0\,\mathrm{nm}$&\multicolumn{2}{c}{$-0.5\,\mathrm{nm/s}$}\\
		\hline	
		Parameter& $\aon$ &$\aoff$&$w_\mathrm{c}$\\
		Value& $3\,\mathrm{nm}$&$0\,\mathrm{nm}$&$0.1\,\mathrm{nm}$\\
		\hline		
		Parameter & $\Vset$&$\Vcond$&$\Vclear$&$\Vread$&$\RG$&$T$\\
		Value& $1.0\,\mathrm{V}$&$0.9\,\mathrm{V}$& $-1.0\,\mathrm{V}$&$0.1\,\mathrm{V}$&$40\,\mathrm{k\Omega}$&$15\,\mathrm{\mu s}$\\			
		\hline\hline
	\end{tabular}
\end{table}

\Cref{eqn:Rmin,eqn:s_ui} are evaluated in order to get the operation constraints imposed by the circuit. Namely, $R_{\mathrm{min},Q} = 101.449\,\mathrm{k\Omega}$ and $s_{\mathrm{min},Q} = 0.908$. We can see that, in Case 1 and assuming no state drift in $P$, the output memristor $Q$ can never reach a state higher than $s_{\mathrm{min},Q}$ or, equivalently, can never have a resistance lower than $R_{\mathrm{min},Q}$.

\subsection{Methodology \& Setup}\label{subsec:method}
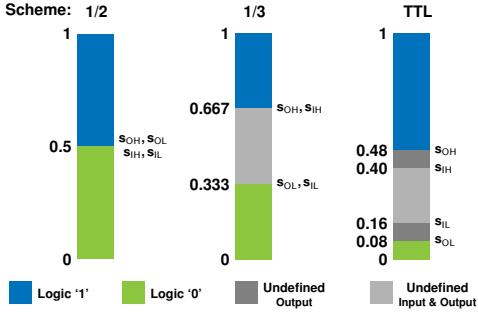
\begin{figure}[t]
	\centering
	\tikzset{%
		Hstyle/.style={fill,RoyalBlue},%
		Lstyle/.style={fill,LimeGreen},%
		lbl/.style={left,font=\sffamily\bfseries\color{black}},%
		logic/.style={left,font=\sffamily\bfseries\color{black}\footnotesize},%
		thresh/.style={logic,right,align=center},
		Indef/.style={fill,gray!60},%
		Ondef/.style={fill,gray}
	}
	\def\bwidth{.8}
	\begin{tikzpicture}[scale=.6, transform shape]
	\node[lbl] at (0,5.55) {Scheme:};
	\begin{scope}[shift={(0,0)}]
	\node[lbl] at (\bwidth,5.5) {1/2};
	\fill[Hstyle] (0,2.5) rectangle (\bwidth,5);
	\fill[Lstyle] (0,0) rectangle (\bwidth,2.5);
	\path[] (0,0) node[lbl]{0}			
	(0,.5*5) node[lbl]{0.5}
	(\bwidth,.5*5) node[thresh] {\footnotesize s$_\mathbf{\mathsf{OH}}$,\,s$_\mathbf{\mathsf{OL}}$\\ s$_\mathbf{\mathsf{IH}}$,\,s$_\mathbf{\mathsf{IL}}$}
	(0,5) node[lbl]{1};
	\end{scope}
	\begin{scope}[shift={(3.5,0)}]
	\node[lbl] at (\bwidth,5.5) {1/3};
	\fill[Hstyle] (0,.667*5) rectangle (\bwidth,5);
	\fill[Lstyle] (0,0) rectangle (\bwidth,.333*5);
	\fill[Indef] (0,.333*5) rectangle (\bwidth,.667*5);
	\path[] (0,0) node[lbl]{0}
	(0,.333*5) node[lbl]{0.333}				
	(\bwidth,.333*5) node[thresh] {s$_\mathbf{\mathsf{OL}}$,\,s$_\mathbf{\mathsf{IL}}$}
	(0,.667*5) node[lbl]{0.667}
	(\bwidth,.667*5) node[thresh] {s$_\mathbf{\mathsf{OH}}$,\,s$_\mathbf{\mathsf{IH}}$}
	(0,5) node[lbl]{1};
	\end{scope}
	\begin{scope}[shift={(7,0)}]
	\node[lbl] at (\bwidth+.2,5.5) {TTL};
	\fill[Hstyle] (0,2.4) rectangle (\bwidth,5);
	\fill[Lstyle] (0,0) rectangle (\bwidth,.4);
	\fill[Ondef] (0,.4) rectangle (\bwidth,2.4);
	\fill[Indef] (0,.8) rectangle (\bwidth,2.0);
	\path[] (0,0) node[lbl]{0}
	(0,.4) node[lbl]{0.08}				
	(\bwidth,.4) node[thresh] {s$_\mathbf{\mathsf{OL}}$}
	(0,.8) node[lbl]{0.16}
	(\bwidth,.8) node[thresh] {s$_\mathbf{\mathsf{IL}}$}
	(0,2) node[lbl]{0.40}
	(\bwidth,2) node[thresh] {s$_\mathbf{\mathsf{IH}}$}
	(0,2.4) node[lbl]{0.48}
	(\bwidth,2.4) node[thresh] {s$_\mathbf{\mathsf{OH}}$}
	(0,5) node[lbl]{1};
	\end{scope}
	\begin{scope}[shift={(-1.5,-1)}]
	\fill[Hstyle] (0,0) rectangle (.5,.5);
	\fill[Lstyle] (2.5,0) rectangle (3,.5);
	\fill[Ondef] (5,0) rectangle (5.5,.5);
	\fill[Indef] (8,0) rectangle (8.5,.5);
	\node[thresh] at (.5,.2) {Logic `1'};
	\node[thresh] at (3,.2) {Logic `0'};
	\node[thresh] at (5.5,.2) {Undefined\\\scriptsize Output};
	\node[thresh] at (8.5,.2) {Undefined\\\scriptsize Input \& Output};
	\end{scope}
	\end{tikzpicture}
	\caption{Different logic thresholds used in this paper.}
	\label{tab:logic_thresholds} 
\end{figure}

\begin{figure}[t]
	\centering
	\includegraphics[width=.25\textwidth]{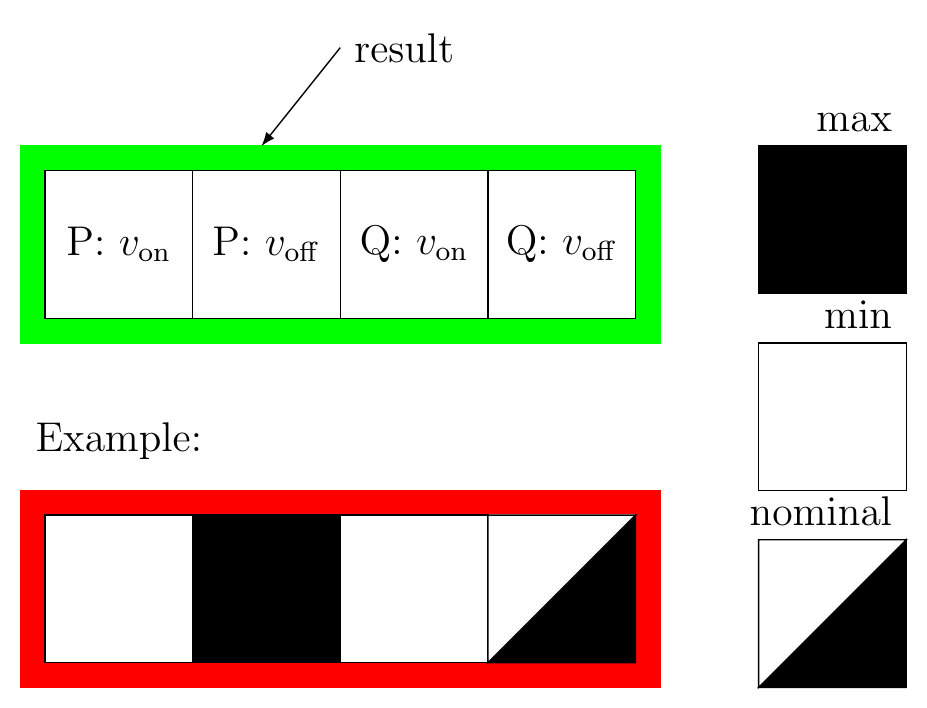}
	\caption{Four squares show the state of each variable in a simulation set and the outline color (green or red) shows the simulation result (correct or failed, respectively).}
	\label{fig:legend_results_variation_von_voff}
\end{figure}
\begin{figure*}[t]
	\centering
	\includegraphics[width=.85\textwidth]{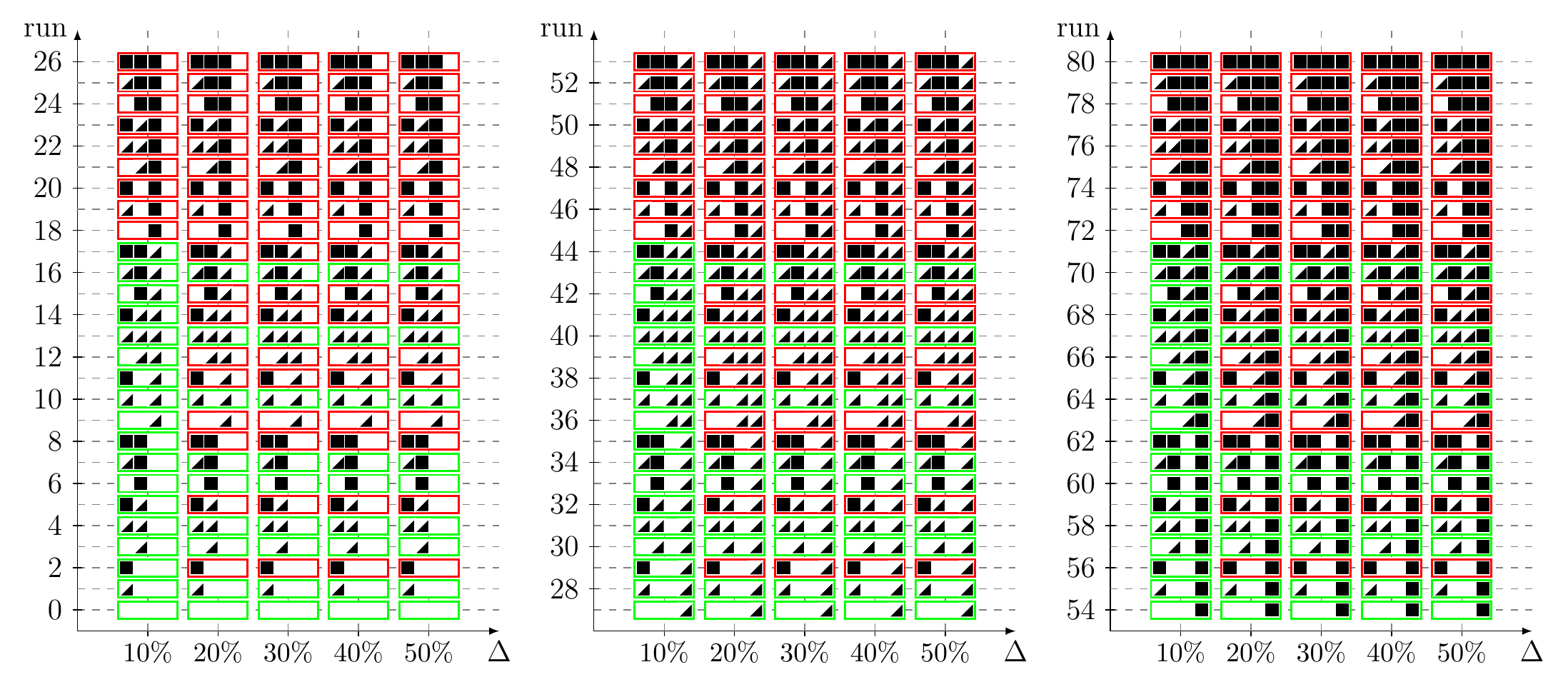}
	\caption{Results summary for different degrees of variation in $\von, \voff$ of $P$ and $Q$. The $1/3$ logic thresholds scheme was used here.} 
	\label{fig:full_results_variation_von_voff}
\end{figure*}

Proper \gls{imply} operation results -- with respect to the output logic thresholds -- are used to determine reliability. Correct operation is ensured when state changes within the memristors are occurring (switching conditions met) and are fast enough to exceed the given logic thresholds. We apply three different logic threshold schemes (shown in \Cref{tab:logic_thresholds}) to evaluate the operation results in relation to the chosen logic threshold. Each scheme defines separate, normalized input ($s_\mathrm{IH}$, $s_\mathrm{IL}$) and output ($s_\mathrm{OH}$, $s_\mathrm{OL}$) thresholds, as in conventional digital logic. Whereas the ``1/2'' and ``1/3'' scheme were chosen arbitrarily, the ``TTL'' scheme is derived from standard TTL ($V_\mathrm{CC}=5\,\mathrm{V}$)~\cite{TI_ttl_thresholds}. This is done by normalizing the threshold voltages $V_\mathrm{IH}$, $V_\mathrm{IL}$, $V_\mathrm{OH}$ and $V_\mathrm{OL}$ to $V_\mathrm{CC}$ -- e.g. $s_\mathrm{IH}=V_\mathrm{IH}/V_\mathrm{CC}$. The range between high and low thresholds, $[s_{\mathrm{IL}},s_{\mathrm{IH}}]$ and $[s_{\mathrm{OL}},s_{\mathrm{OH}}]$, is forbidden; in other words, the logic values and states in those ranges are considered undefined. 
Reasons for failures are not separately determined in our setup. Hence, failures during initialization, which lead to erroneous operation results, are counted as regular failures and are not distinguished from errors during the operation itself. Further, our simulation setup utilizes constant timesteps, so actual switching time are not explicitly measured. \par 

To obtain a nominal timebase for the \gls{imply} gate, a transient analog simulation of the memristor model was conducted. Examining the resulting waveform of the normalized state $s$ after the simulation showed that it takes $15\,\mathrm{\mu s}$ to switch from $1\%$ to $99\%$ of the state boundaries. Thus, the timestep of circuit operation is set to $T=15\,\mathrm{\mu s}$. Every action (initialization, \gls{imply} operation, readout) is executed using this fixed timestep. \par 

Analog transient simulations were conducted in LTSpice, making use of this setup. Two memristor parameters per device ($\Ron, \Roff$ or $\von, \voff$ or $\kon, \koff$) were varied simulatenously within the ranges reported in measurements~\cite{CAS_physRealization} and relative to the nominal state with a maximum deviation of $\pm 50\%$. 

\subsection{Result Presentation Method}\label{subsec:postprocess}
To display the numerous results, we have come up with a presentation method of our own, which we introduce here. \par 

Each parameter set is represented by a group of four squares. The left two squares, as displayed in \Cref{fig:legend_results_variation_von_voff}, show parameter values of memristor $P$, and the right two show that of memristor $Q$. The filling of each square represents the state of the corresponding parameter: empty means minimum, half-filled nominal and fully filled 
maximum. \Cref{fig:legend_results_variation_von_voff} shows this concept and provides an example, too. The outline color of the squares shows whether the simulation result for a set of parameter variation  ($\Delta$) was correct (highlighted by green) or incorrect (highlighted by red). In general, any combination of four parameters can be varied concurrently and displayed this way. However, our approach was to use three parameter sets: $\{\Ronp, \Roffp, \Ronq,\Roffq\}$, $\{\vonp,\voffp,\vonq,\voffq\}$ and $\{\konp,\koffp,\konq,\koffq\}$, as explained in \Cref{subsec:method}. \Cref{fig:full_results_variation_von_voff} shows a {complete set of simulations} for the parameters $\{\vonp,\voffp,\vonq,\voffq\}$. These resulting sets are then used to quickly identify those parameters that are common between different failed runs. For example, \Cref{fig:full_results_variation_von_voff} shows that the \gls{imply} operation produces no correct output if either, $\vonq$ or $\vonp$, is at its maximum value for variations greater than or equal to $10\%$.

\subsection{Results analysis}\label{subsec:sim_analysis}
Combining the math provided in \Cref{sec:math} and the simulation results obtained in \Cref{subsec:postprocess} into joint graphs gives us \Cref{fig:R_P_limit_m2_v,fig:R_P_limit_m2_R,fig:R_P_limit_m2_k,fig:R_P_limit_m0_v,fig:R_P_limit_m1_v}. First, we take a closer look at \Cref{fig:R_P_limit_m2_v,fig:R_P_limit_m2_R,fig:R_P_limit_m2_k}, because they represent the most relatable logic threshold scheme, derived from traditional TTL thresholds. \Cref{fig:R_P_limit_m0_v,fig:R_P_limit_m1_v} show the same equations as in \Cref{fig:R_P_limit_m2_v}, plotted for the 1/2 and 1/3 threshold scheme, respectively. The other two graphs for these logic threshold schemes are omitted  as they lead to the same conclusions as \Cref{fig:R_P_limit_m0_v,fig:R_P_limit_m1_v}. Furthermore, the threshold voltages turned out to be the most critical parameters, so special attention is given to their results.

\subsubsection{Graph structure}\label{subsubsec:graph_structure}
Here, we explain how these graphs are composed. 
Parameters $\Ronp$ and $\Roffp$ of memristor $P$ are always shown on the y-axis since $R_P$ is crucial for the outcome of the operation. 
We can also see that from the fact that $\Roffp$ or $\Ronp$ are present in all of the constraints described in \Cref{sec:math}. Different parameters are used in each graph for the x-axes. \par 

Colored curves and areas are used to show constraints and important ranges:
\newif\ifgrayscale\grayscalefalse
\definecolor{SandRed}{HTML}{8c6b6b}
\definecolor{SandBlue}{HTML}{5a7184}
\definecolor{SandGreen}{HTML}{76a290}
\definecolor{SandOrange}{HTML}{ebb134}
\definecolor{SandPurple}{HTML}{B57DA5}
\grayscalefalse
\def\barfr{none}
\ifgrayscale
	\colorlet{RonAreaColor}{black!10}
	\colorlet{RonHatchColor}{black}
	\colorlet{RoffVqAreaColor}{black!50}
	\colorlet{RoffHatchColor}{black}
	\colorlet{RoffVpAreaColor}{black!30}
	\colorlet{OkColor}{white}
	\colorlet{FailColor}{black}
	\colorlet{UnsureColor}{black!50}
	\colorlet{OtherErrorColor}{black!90}
	\colorlet{StaticRColor}{black!80}
	\colorlet{StaticvColor}{black!30}
	\colorlet{StaticRonColor}{black!80}
	\colorlet{StaticvRonColor}{black!30}
	\colorlet{DynVqColor}{black!60}
	\colorlet{DynVpRoughColor}{black!40}
	\colorlet{DynVpArMeanColor}{black!20}
	\colorlet{DynVpGeoMeanColor}{black!30}
	\colorlet{ThreshColor}{black}
	\colorlet{ZerovColor}{white}
	\colorlet{RoffAreaColor}{black!50}
	\colorlet{DynRoffqColor}{black!60}
	\colorlet{DynRonqColor}{black!40}
	\colorlet{ThreshpColor}{black}
	\colorlet{ThreshqColor}{black!30}
	\colorlet{RoffKqAreaColor}{black!50}
	\colorlet{RoffKpAreaColor}{black!30}
	\colorlet{DynKqColor}{black!60}
	\colorlet{DynKpRoughColor}{black!40}
	\colorlet{DynKpArMeanColor}{black!20}
	\colorlet{DynKpGeoMeanColor}{black!30}
	\def\barfr{black}	
\else
	\colorlet{RonAreaColor}{SandPurple!60}
	\colorlet{RonHatchColor}{SandPurple}
	\colorlet{RoffVqAreaColor}{white}
	\colorlet{RoffHatchColor}{SandBlue!45}
	\colorlet{RoffVpAreaColor}{white}
	\colorlet{OkColor}{green!70}
	\colorlet{FailColor}{red!80}
	\colorlet{UnsureColor}{Orange!50!Yellow}
	\colorlet{OtherErrorColor}{FailColor}
	\colorlet{StaticRColor}{ForestGreen}
	\colorlet{StaticvColor}{Orange!90!black}
	\colorlet{StaticRonColor}{RedViolet}
	\colorlet{StaticvRonColor}{Black!70}
	\colorlet{DynVqColor}{blue!50!SandBlue}
	\colorlet{DynVpRoughColor}{Sepia}
	\colorlet{DynVpArMeanColor}{lime}
	\colorlet{DynVpGeoMeanColor}{OliveGreen}
	\colorlet{ThreshColor}{Cyan}
	\colorlet{ZerovColor}{RoyalBlue}
	\colorlet{RoffAreaColor}{SandBlue!60}
	\colorlet{DynRoffqColor}{StaticvColor}
	\colorlet{DynRonqColor}{blue!50!SandBlue}
	\colorlet{ThreshpColor}{Cyan}
	\colorlet{ThreshqColor}{LimeGreen}
	\colorlet{RoffKqAreaColor}{white}
	\colorlet{RoffKpAreaColor}{white}
	\colorlet{DynKqColor}{blue!50!SandBlue}
	\colorlet{DynKpRoughColor}{Sepia}
	\colorlet{DynKpArMeanColor}{lime}
	\colorlet{DynKpGeoMeanColor}{OliveGreen}
	\def\barfr{none}
\fi
\def\itlen{5mm} 
\def\itsep{1.5mm} 
\def\itdum{-.8ex} 
\begin{itemize}
	\item[%
		\tikz{\draw[black,thin,dashed] (0mm,0mm)--(\itlen,0mm);%
			  \path (0mm,\itdum)--(\itlen,\itdum);}]%
		Black, dashed lines indicate nominal parameter values
	\item[%
		\tikz{\draw[ThreshColor, thick] (0mm,0mm)--(\itlen,0mm);%
			  \draw[ThreshColor,thick,dashed](0mm,-\itsep)--(\itlen,-\itsep);}]%
		Light blue lines show input logic thresholds $R_\mathrm{IH}$ (solid) and $R_\mathrm{IL}$ (dashed)
		for memristor $P$. 
	\item[%
		\tikz{\draw[StaticRonColor, thick] (0mm,0mm)--(\itlen,0mm);%
		  	  \draw[DynVqColor, thick] (0mm,-.5*\itsep)--(.5*\itlen,-.5*\itsep);%
		  	  \draw[DynVqColor, thick,dotted] (.5*\itlen,-.5*\itsep)--(\itlen,-.5*\itsep);%
		  	  \draw[DynVpArMeanColor, thick] (0mm,-\itsep)--(.5*\itlen,-\itsep);%
		  	  \draw[DynVpArMeanColor, thick,dotted] (.5*\itlen,-\itsep)--(\itlen,-\itsep);}]%
	  	 Colored curves show the constraints from \Cref{sec:math}. Dotted parts indicate invalid plotting ranges, which do not correspond to any real value in physical devices. 
	\item[%
		\tikz{\draw[StaticRColor, thick,-latex] (0mm,.5*\itsep)--(\itlen,.5*\itsep);%
		  	  \draw[DynVqColor, thick,latex-] (0mm,-.5*\itsep)--(\itlen,-.5*\itsep);}]%
		Arrows indicate how the constraints restrict the operating area of a parameter, i.e., which side of the curve is acceptable due to the given constraint.
	\item[%
		\tikz{\fill[RoffHatchColor] (0mm,0mm) rectangle (\itlen,-\itsep);}]%
	    Blue areas show valid ranges of $\Roffp$ and the respective parameters on the x-axes. For example, in \Cref{fig:R_P_limit_m2_v}, this area represents valid ranges of $\Roffp$ versus $\vonq$, $\voffq$, $\vonp$ and $\voffp$. Note that for $\vonp$ our recommended range was used to limit the valid area, as the three different curves are only weak constraints.
	\item[%
		\tikz{\fill[RonAreaColor] (0mm,0mm) rectangle (\itlen,-\itsep);}]%
		Purple areas show valid ranges of $\Ronp$ and the respective parameters on the x-axes. For example, in \Cref{fig:R_P_limit_m2_k}, this area represents valid ranges of $\Ronp$ versus $\konq$, $\koffq$, $\konp$ and $\koffp$. Note that restrictions on x-axis parameters are inherited from the $\Roffp$ operating area.
	\item[%
		\tikz{\draw[FailColor,line width=1mm] (0mm,0mm)--(.33*\itlen,0mm);%
			  \draw[UnsureColor,line width=1mm](.33*\itlen,0mm)--(.66*\itlen,0mm);%
		  	  \draw[OkColor,line width=1mm] (.66*\itlen,0mm)--(\itlen,0mm);%
	  	  	  \path(0mm,\itdum)--(\itlen,\itdum);}]%
	  	The bars at each side of the graphs overlay our simulation results. Red sections show incorrect \gls{imply} results, green sections show correct results and orange sections are used for ranges in between, which are not explicitly covered by the simulations. 
\end{itemize}

\subsubsection{Variation in voltage threshold}\label{par:v_var}

\begin{figure}
	\centering
	\includegraphics[width=\linewidth]{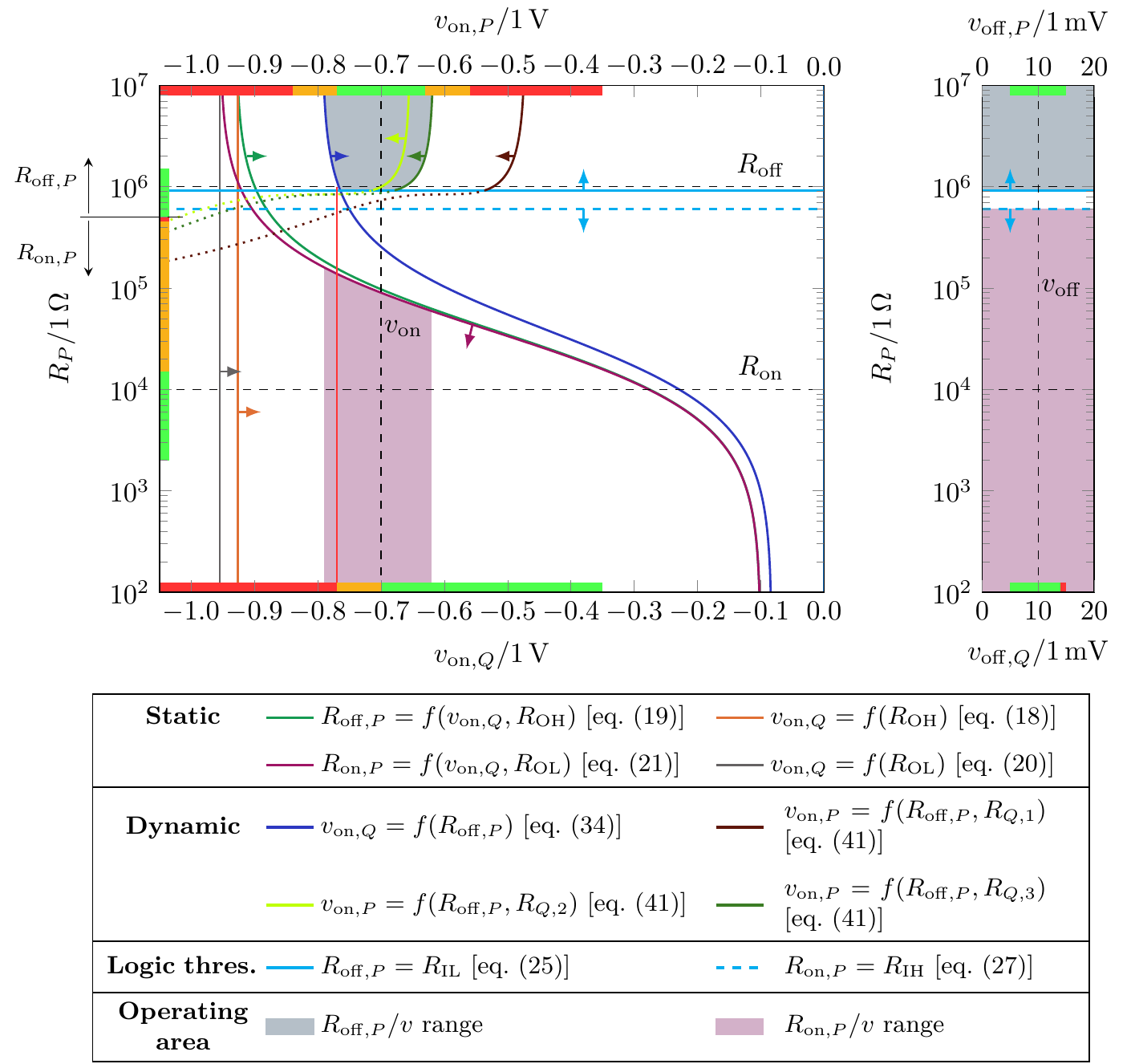}
	\caption{Analytical constraints and logic thresholds for the TTL scheme plotted over a range of memristor parameters $\{\Rp,v_P, v_Q\}$. }
	\label{fig:R_P_limit_m2_v}
\end{figure}

\begin{figure}
	\centering
	\includegraphics[width=\linewidth]{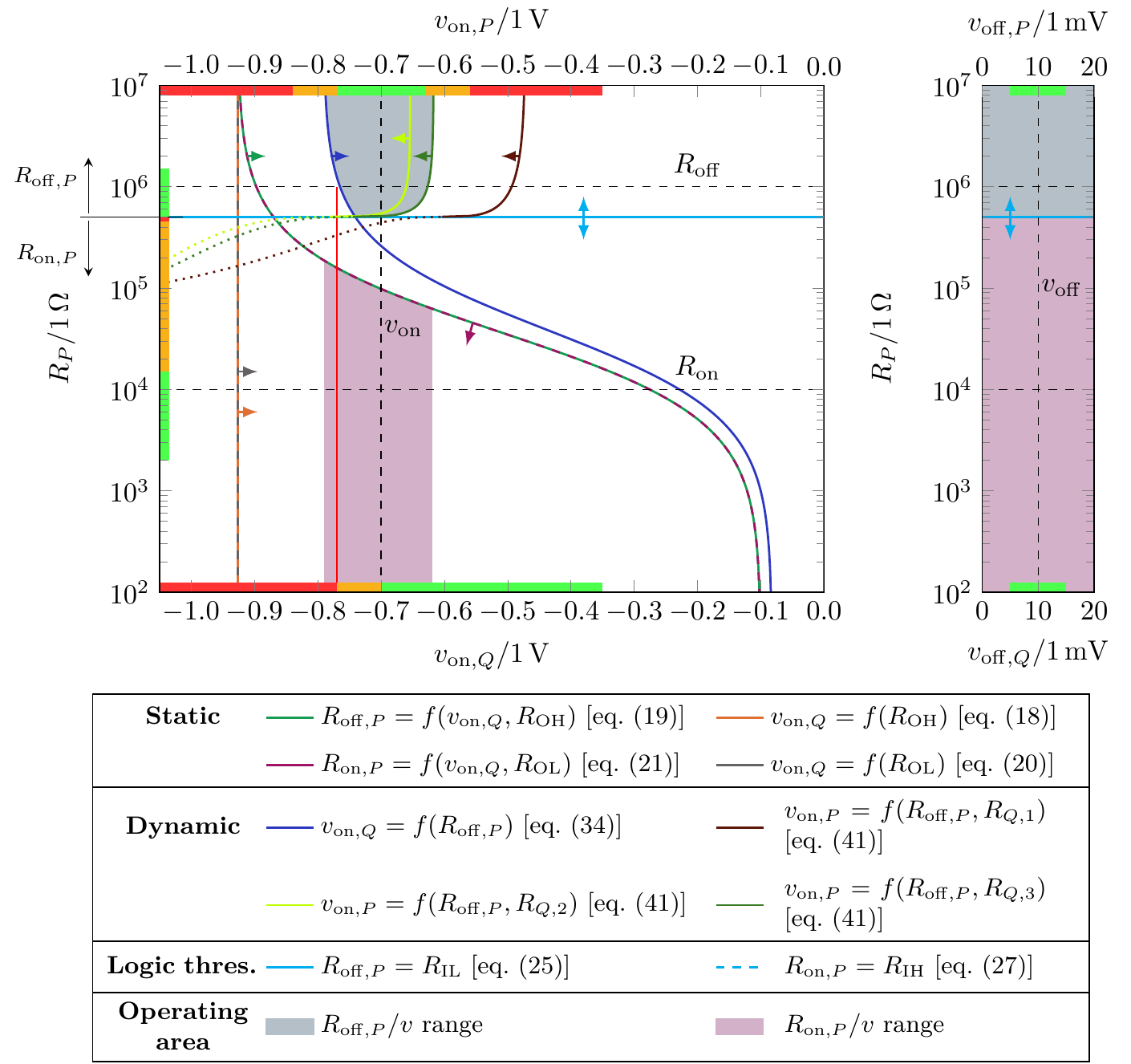}
	\caption{Analytical constraints and logic thresholds for the 1/2 scheme plotted over a range of memristor parameters $\{\Rp,v_P, v_Q\}$. }
	\label{fig:R_P_limit_m0_v}
\end{figure}

\begin{figure}
	\centering
	\includegraphics[width=\linewidth]{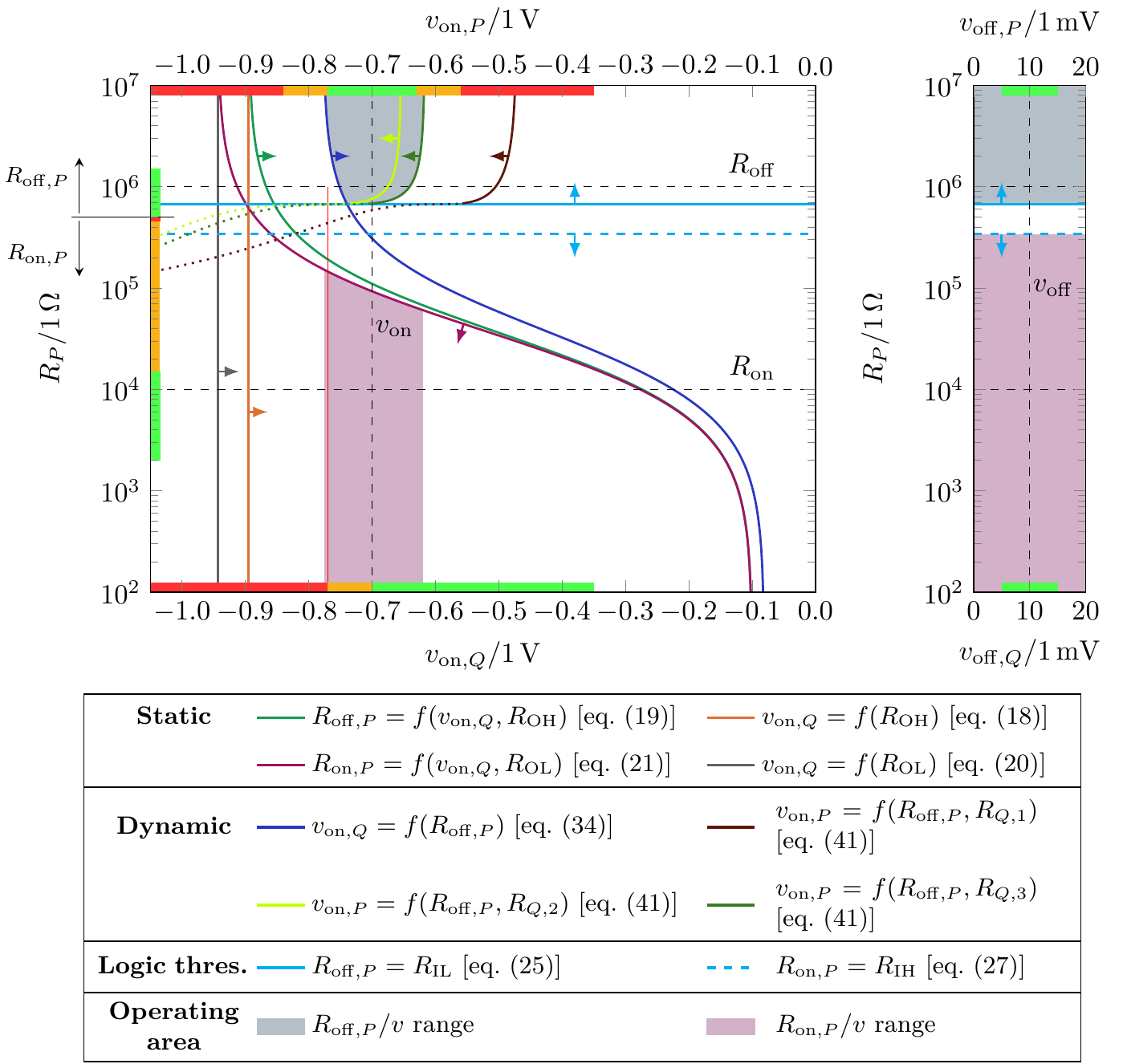}
	\caption{Analytical constraints and logic thresholds for the 1/3 scheme plotted over a range of memristor parameters $\{\Rp,v_P, v_Q\}$.} 
	\label{fig:R_P_limit_m1_v}
\end{figure}

\Cref{fig:R_P_limit_m2_v,fig:R_P_limit_m0_v,fig:R_P_limit_m1_v} depict voltage thresholds $\vonp$, $\voffp$, $\vonq$ and $\voffq$ of memristor $P$ and $Q$, as well as resistance parameters $\Ronp$ and $\Roffp$ of $P$ using different logic threshold schemes (\Cref{tab:logic_thresholds}). For the analysis we concentrate on the TTL scheme, \Cref{fig:R_P_limit_m2_v}. \par 

The logic thresholds ($R_\mathrm{IH}$, $R_\mathrm{IL}$) divide the plot into two parts: The bottom part concerning $\Ronp$ and the top part concerning $\Roffp$. Adding the static constraints, \Cref{eqn:vonq_case1,eqn:vonq_case3,eqn:Roffp_case1,eqn:Ronp_case3}, on top of the logic thresholds decreases the valid range of $\Roffp$, $\vonq$ and in particular $\Ronp$. The latter is evident from the purple area in \Cref{fig:R_P_limit_m2_v}, which is smaller than the plotted range. However, regarding $\Roffp$ and $\vonq$, the dynamic constraint, \Cref{eqn:vonq_wmin}, is even stricter than the static constraint. \par 

There are no static constraints for $\vonp$. A rough dynamic estimation is provided by \Cref{eqn:vonp_wmax}, which depends on $V_{P\mathrm{f}}$.  As discussed in \Cref{subsec:dyn_math}, \Cref{eqn:vonp_wmax} is evaluated three times, using $R_{Q,1}$, $R_{Q,2}$ and $R_{Q,3}$, respectively. The three curves are drawn in brown, dark green and light green. No constraint for $\voff$ has been found (\Cref{sec:math}). Hence, the valid ranges of $R_P$ over $\{\voffp, \voffq\}$ are only limited by logic thresholds, \Cref{eqn:Roffp_greater_RIL,eqn:Ronp_smaller_RIH}. As a consequence of the above constraints, the valid range for each parameter is decreased and thus the advisable operating area remains as shown by the colored areas. 

Simulation results for variation in $\vonq$ show very good agreement with the mathematical analysis, especially the dynamic estimation in \Cref{eqn:vonq_wmin}, which depends on \Cref{eqn:VQi,eqn:wmin}. At $+10\%$ variation of $\vonq$ and nominal $\Roffp$, the simulation fails (indicated by the thin red line), as the analysis predicted. \Cref{fig:R_P_limit_m2_v} shows very clearly that this failure is not accurately predicted by the static constraints from \Cref{subsec:stat_math} alone. Hence, the dynamic estimation (\Cref{subsec:dyn_math}) is vital. Variation in $\vonp$ strengthens this point further, since different methods of estimating the dynamic behavior leads to important changes regarding the agreement of the simulations and the derived analytical constraints. On the upper end of the $\vonp$ range, \Cref{eqn:vonp_wmax} ({evaluated using $R_Q=R_{Q,3}$ for $V_{P\mathrm{f}}$, \Cref{eqn:vpf}}) provides good congruence with our simulations, whereas \Cref{eqn:vonp_wmax} (evaluated using $R_Q=R_{Q,2}$ for $V_{P\mathrm{f}}$, \Cref{eqn:vpf}) represents a more conservative estimation.  In contrast, evaluating \Cref{eqn:vonp_wmax} using the theoretical minimum $R_Q=R_{Q,1}=R_{\mathrm{min},Q}$ in \Cref{eqn:vpf}, does not yield a good estimation. On the lower end of the $\vonp$ range, simulation results indicate some failures for $\vonp\leq -0.84\,\mathrm{V}$ ($+20\%$). This behavior cannot be explained by any of the constraints from \Cref{sec:math}. 
According to the simulation results (\Cref{subsec:postprocess}, \Cref{fig:full_results_variation_von_voff}), these specific failures only occur when $\vonq\geq -0.7\,\mathrm{V} (\pm 0\%)$, which leads us to believe that the reason for failure is the $20\%$ mismatch between $\vonp$ and $\vonq$. Regarding both, $\voffp$ and $\voffq$, there are almost no failures as expected, except for a (minor) failure during initialization for $\voffq$ at $+50\%$.\par 

In terms of $R_P$ variation, the simulation results suggest that $\Roffp$ can lie within the uncertain range between logic thresholds while the \gls{imply} operation still outputs correct results. This stands to reason since the thresholds are artificial limits not governed by the circuit behavior. Further, $\Ronp$ is fine up to the lowest simulated value of $\Roffp$, because at that point $\Roffp>\Ronp$ changes to $\Roffp<\Ronp$, and hence the operation fails. \par 

Combining all the simulation results and their respective analytical constraints, we can identify the areas in which the circuit is most likely to operate correctly. These are the areas highlighted in \Cref{fig:R_P_limit_m2_v,fig:R_P_limit_m0_v,fig:R_P_limit_m1_v}. \Cref{eqn:vonq_wmin,eqn:vonp_wmax} and their respective dependencies, \Cref{eqn:VQi,eqn:wmin,eqn:vpf} (evaluated using $R_Q=R_{Q,3}$),  are recommended for estimating the valid ranges of $\Roffp$ versus $\{\vonp,\vonq\}$; whereas the static constraints \green{\Cref{eqn:vonq_case1,eqn:vonq_case3,eqn:Roffp_case1,eqn:Ronp_case3}} are sufficient for $\Ronp$ versus $\{\vonp,\vonq\}$.

\begin{figure}[t]
	\centering
	\includegraphics[width=\linewidth]{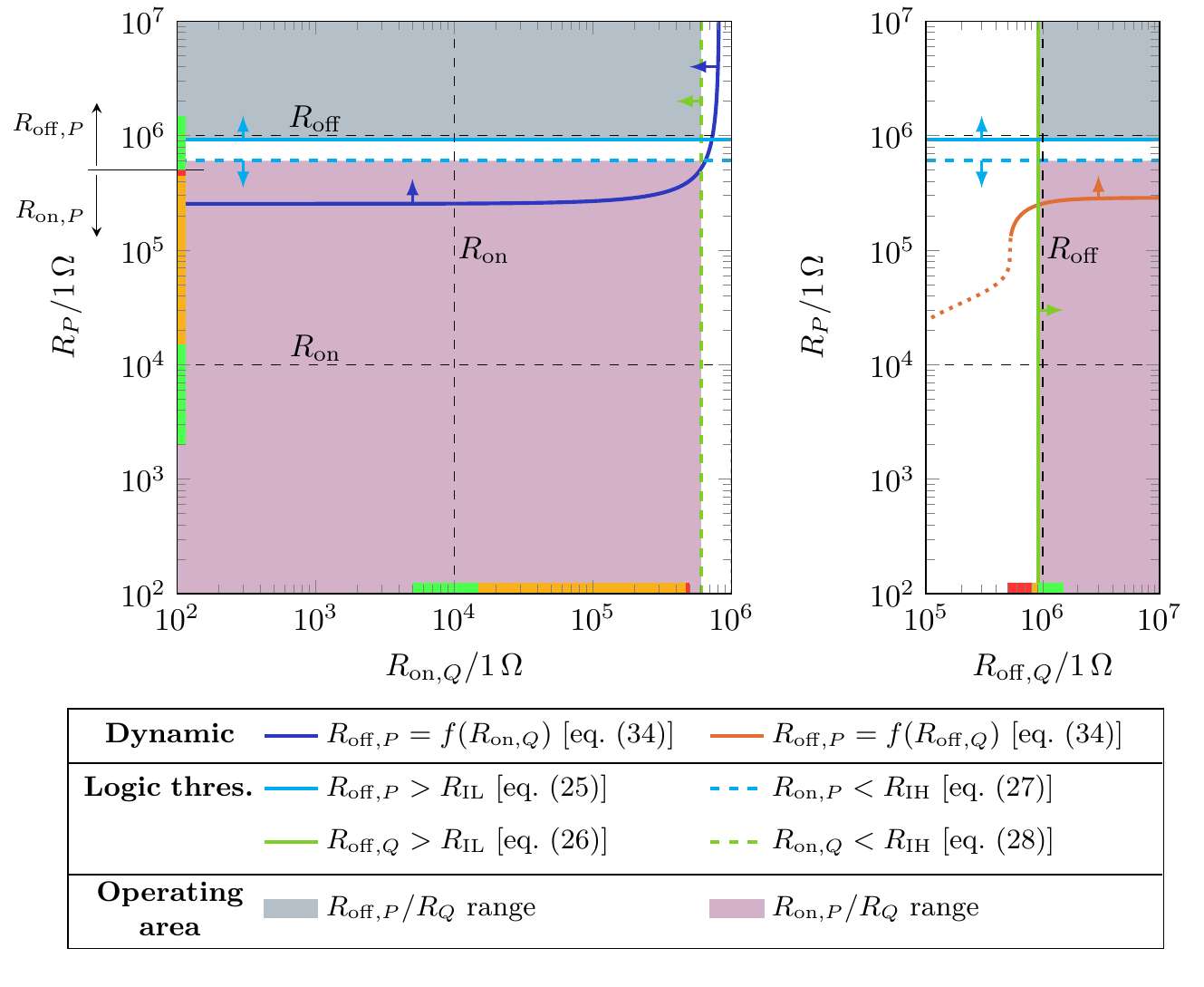}		
	\caption{Analytical constraints and logic thresholds for the TTL scheme plotted over a range of memristor parameters $\{\Rp, \Rq\}$.} 
	\label{fig:R_P_limit_m2_R}
\end{figure}

\subsubsection{Variation in resistance limits}\label{par:r_var}
There are no static constraints limiting $\Ronq$ or $\Roffq$. Therefore, only logic thresholds and the dynamic estimation of \Cref{eqn:vonq_wmin} can be applied. The latter depends on \Cref{eqn:VQi,eqn:wmin} and is evaluated in two ways: First, varying $\Ronq$, but not $\Roffq$; and second varying $\Roffq$, but not $\Ronq$. It is interesting to see that -- for any of the three schemes of \Cref{tab:logic_thresholds} -- the logic thresholds limit the operating areas (blue and purple) much more than the actual analytical constraints. Simulation results for $\Ronp$ and $\Roffp$ are identical to \Cref{fig:R_P_limit_m2_v}, however, $\Roffq$ cannot reach as low as $\Roffp$ without causing a failure.  This is solely due to the chosen logic thresholds, as an \gls{imply} output of $\Roffq<R_\mathrm{OL}$ is considered as failure. \par 

Overall, resistance variation does not seem to hold as much potential for failures as variation in threshold voltage(s) does.  \Cref{eqn:vonq_wmin} and its dependencies, \Cref{eqn:VQi,eqn:wmin}, can be used to identify valid parameter ranges, but -- based on our simulation results -- it is most likely not necessary. This is true for all three logic threshold schemes listed in \Cref{tab:logic_thresholds}. 

\begin{figure}
	\centering
	\includegraphics[width=\linewidth]{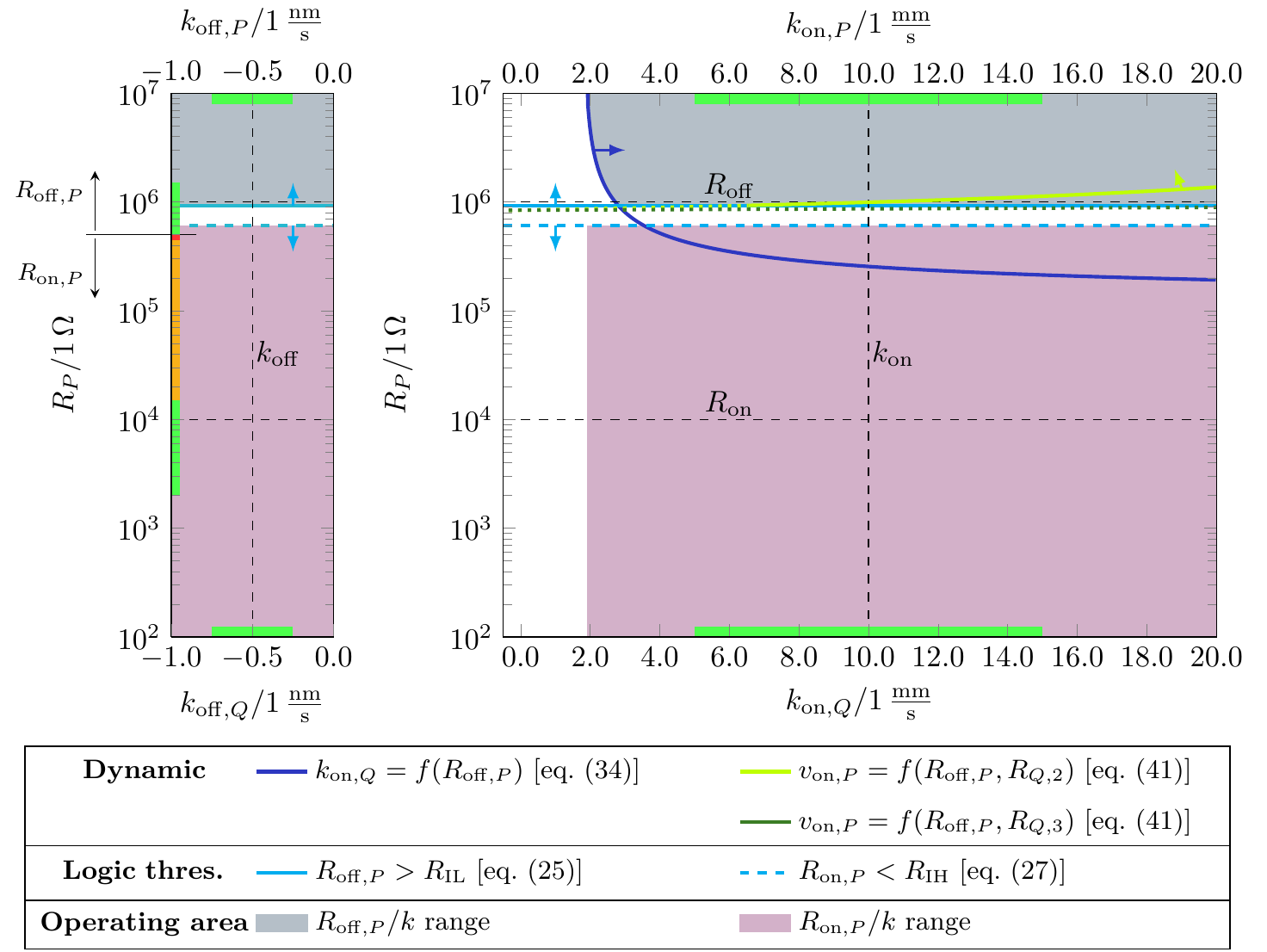}
	\caption{Analytical constraints and logic thresholds for the TTL scheme plotted over a range of memristor parameters $\{\Rp,k_P,k_Q\}$. }
	\label{fig:R_P_limit_m2_k}
\end{figure}

\subsubsection{Variation in switching speed}\label{par:k_var}
The dynamic constraint in \Cref{eqn:vonq_wmin} can be used to extract limits of $\konq$, while \Cref{eqn:vonp_wmax} provides the basis for the analysis of $\konp$. \Cref{fig:R_P_limit_m2_k} shows the plotted equations and logic thresholds. \Cref{eqn:vonp_wmax} (evaluated using \Cref{eqn:vpf}, where $R_Q=R_{Q,1}$) is omitted, as well as all constraints containing $\koffp$ and $\koffq$, since they are far outside of the plotted range.
The graph in \Cref{fig:R_P_limit_m2_k} shows that $\konp$ is hardly restricted by any constraint. Only at relatively high values, greater than $+50\%$ variation, \Cref{eqn:vonp_wmax} (evaluated using \Cref{eqn:vpf} with $R_Q=R_{Q,2}$) comes into effect, but cannot be compared to simulation results, as our simulated range ends at $+50\%$, in compliance with our methodology (\Cref{subsec:method}). In contrast, \Cref{eqn:vonq_wmin} provides a reasonable constraint for $\konq$. Nonetheless, our simulated range only reaches down to $-50\%$ and thus results cannot be compared to the constraint. The other two logic threshold schemes show similar behavior. As before, the colored areas indicate the merged, predicted functional range of both, $\konp$ and $\konq$. 
\par 

In conclusion, switching speed $k$ of both memristors can vary at least by $\pm50\%$ without performance issues, according to our simulation. Analytical constraints suggest that there is a lower boundary for $\konq$ at approximately $2\,\mathrm{mm/s}$ ($-80\%$). 

\section{Simulation -- Crossbar}\label{sec:sim_xbar}
\subsection{Setup}\label{subsec:xbar_circuit}
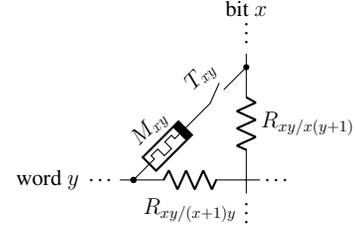
\begin{figure}
	\centering
	\begin{circuitikz}[font=\Large,scale=.6,transform shape]
		\node[rotate=0] at(-.7,0) {\dots};
		\node at (-1.9,0) {word $y$};
		\node[rotate=90] at(2.5,3.3) {\dots};
		\node at (2.5,3.8) {bit $x$};
		\node[rotate=0] at (3.2,0) {\dots};
		\node[rotate=90] at (2.5,-.7) {\dots};
		\draw(2.5,2.8) to[short,-*] (2.5,2.5) to [R,l=$R_{xy/x(y+1)}$] (2.5,0) to[short] (2.5,-.3)
		(-.3,0) to[short,-*] (0,0) to[R,l_=$R_{xy/(x+1)y}$] (2.5,0) to[short](2.8,0);
		\draw (0,0) to[memristor,*-,l=$M_{xy}$] (1.5,1.5) to[short] (1.7,1.7) to[short] (1.9,2.15)  (2,2) to [short](2.5,2.5);
		\node[rotate=45] at (1.5,2.3) {$T_{xy}$};
	\end{circuitikz}
	\caption{Structure of a single cell within the 1T1R crossbar, including line resistances.}
	\label{fig:xbar_cell}
\end{figure}

Analogous to the single \gls{imply} gate simulation setup (\Cref{subsec:sim_circuit}), the circuit in \Cref{fig:imply_gate} is the basis for the crossbar simulation. A complete \xbarsize{} cell 1T1R crossbar circuit was used. The \gls{imply} gate  is formed by two memristors arbitrarily located within the crossbar. Each memristor has its own access device, in our case an ideal switch, and is connected to adjacent cells via resistors that model the nanowire resistances. The ideal switch is modeled using an on-resistance of $1\,\mathrm{\mu\Omega}$ and an off-resistance of $100\,\mathrm{M\Omega}$. Line resistances were chosen to be $10\,\mathrm{\Omega}$ each, according to the worst case in~\cite{sneak_paths_closed_form}. \Cref{fig:xbar_cell} shows the structure of a single cell. 

Circuit parameters of the \gls{imply} gate are identical to \Cref{subsec:sim_circuit}, \Cref{tab:parameters}. Bit-line drivers are attached at the top and bottom for symmetry. The readout strategy  described in \Cref{sec:xbar_fundamental} was implemented. Analog transient simulations were conducted in Cadence Spectre. The method of parameter variation is the same as defined for the single gate in \Cref{subsec:method}, except that only relative parameter variations ($\pm 50\%$) were conducted for the crossbar.

\begin{figure}[b]
	\centering
	\begin{tikzpicture}
	\begin{axis}[
	ybar,
	ymin=0,
	grid=major,
	xtick distance=0.25,
	width=0.6\linewidth,
	height=.35\linewidth,
	xlabel=$s$,
	ylabel={$n$ per bin},
	font=\footnotesize,
	]	
	\addplot +[hist={bins=100}] table [y index=0] {data/init_states.csv};
	\end{axis}
	\end{tikzpicture}\vspace{-3mm}
	\caption{Histogram of initial (normalized) device states, $s$, within the \xbarsize{} crossbar, plotted using 100 bins.}
	\label{fig:xbar_initial_histogram}
\end{figure}
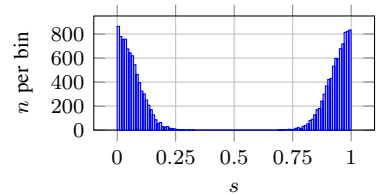

\begin{figure*}
	\centering
	\includegraphics[width=.85\textwidth]{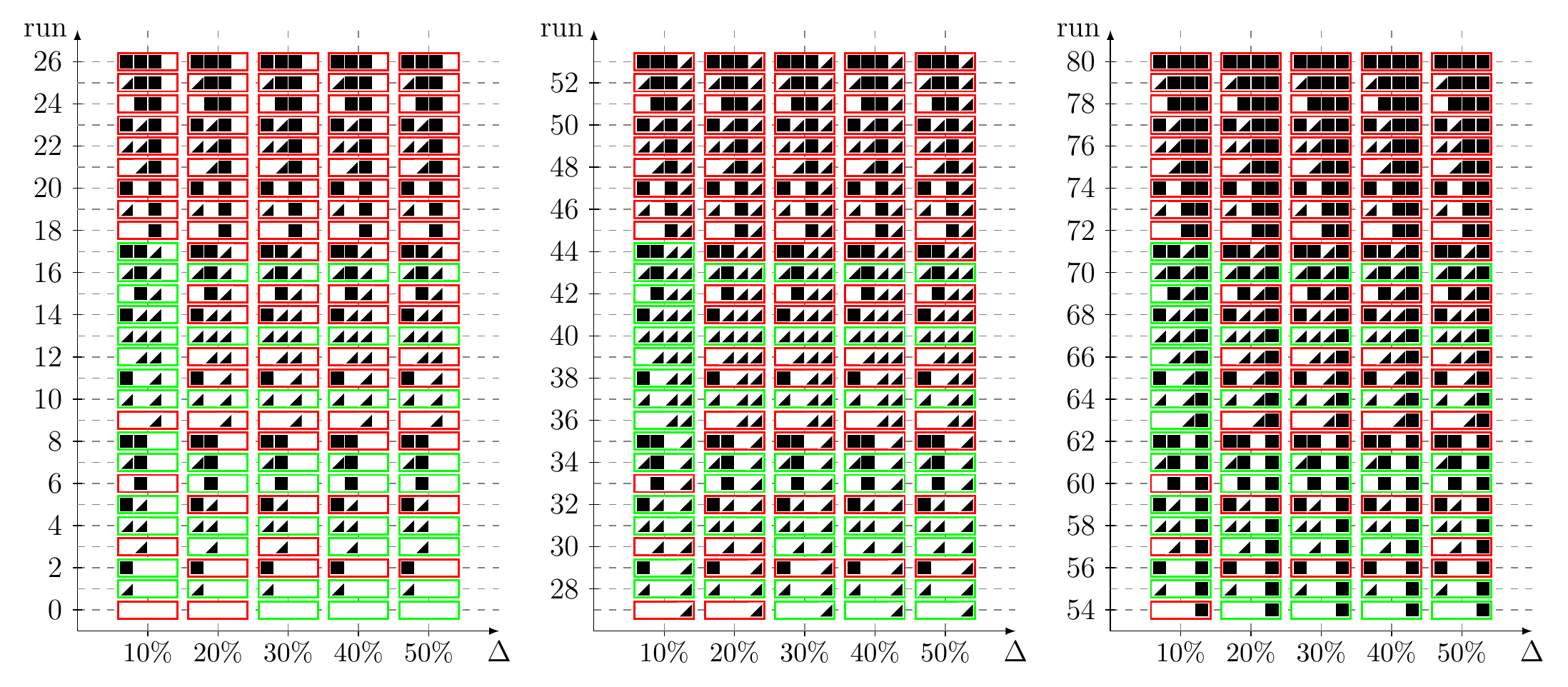}
	\caption{Crossbar results summary for different degrees of variation in $\von, \voff$ of $P$ and $Q$. Logic thresholds for `1' and `0' were set according to the TTL threshold scheme (\Cref{tab:logic_thresholds}). Memristor $P$ was at position $\{0,0\}$, while $Q$ was at the center, $\{63,63\}$.}
	\label{fig:xbar_full_results_variation_von_voff}
\end{figure*}

\subsection{Methodology}\label{subsec:xbar_method}
\gls{imply} gates can be formed by any two memristors in the crossbar. Both, the worst case scenario in terms of parasitic resistance between the two memristors forming a gate, and the worst case voltage drop, were considered. Hence, four separate simulations were conducted with $P$ and $Q$ at different $\{\text{bit},\text{word}\}$ positions. 
\begin{enumerate}
	\item Memristor $P$ at position $\{0,0\}$, $Q$ at position $\{127,127\}$
	\item Memristor $P$ at position $\{127,127\}$, $Q$ at position $\{0,0\}$ 
	\item Memristor $P$ at position $\{0,0\}$, $Q$ at position $\{63,63\}$
	\item Memristor $P$ at position $\{63,63\}$, $Q$ at position $\{0,0\}$
\end{enumerate}

Instead of using idealized ($s=0$ or $s=1$) or manually fixed initial memristor states, each cell is assigned a different initial state during (automated) netlist generation. The states are generated via Octave and follow a Gaussian distribution which has been cut in half as shown in \Cref{fig:xbar_initial_histogram}. Although this approach requires a greater effort, it represents a more realistic scenario than ideal initial states. 

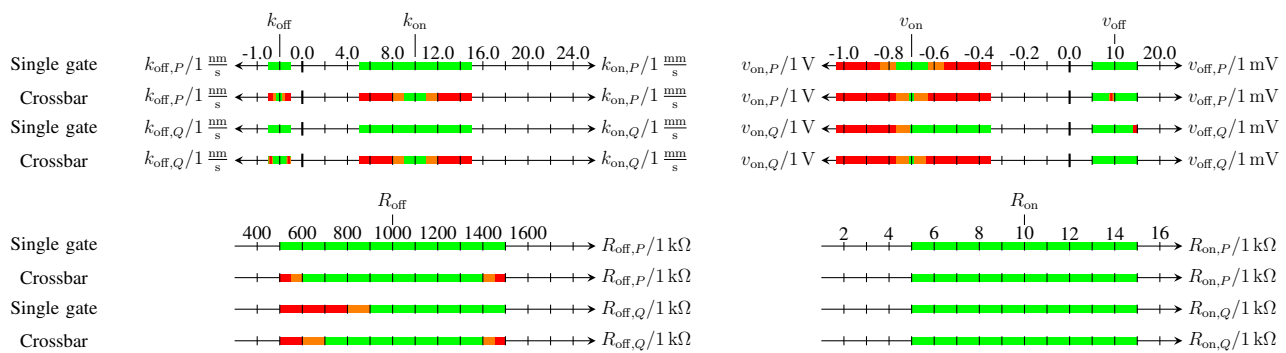
\begin{figure*}[]
	\centering
\def\singletxt{Single gate}
\def\xbartxt{Crossbar}
\def\barsep{-.7} 
\def\groupysep{-4} 
\def\groupxsep{13} 
\begin{tikzpicture}[scale=.6,transform shape,font=\large]
\begin{scope}[shift={(0,0)}]
\node[] at (-8,0) {\singletxt};
\draw[stealth-stealth] (-4,0) node[left]{$\koffp/1\,\mathrm{\frac{nm}{s}}$} -- (4,0)node[right]{$\konp/1\,\mathrm{\frac{mm}{s}}$};
\fill[green] (-1.25,-.08) rectangle (1.25,.08);
\fill[green] (-3.25,-.08) rectangle (-2.75,.08);
\draw[] (-3.0,.1) -- (-3.0,.7) node[above]{\koff};
\draw[] (0,.1) -- (0,.7) node[above]{\kon};
\draw[thick] (-2.5,-.15) -- (-2.5,.15);
\foreach \x/\t in {-3.5/-1.0,-3.0/,-2.5/0.0,-2/,-1.5/4.0,-1/,-.5/8.0,0/,.5/12.0,1/,1.5/16.0,2/,2.5/20.0,3/,3.5/24.0}{
	\draw[] (\x,-.1) -- (\x,.1) node[above,fill=white,inner sep=1pt] {\t};
}
\end{scope}
\begin{scope}[shift={(0,\barsep)}]
\node[] at (-8,0) {\xbartxt};
\draw[stealth-stealth] (-4,0) node[left]{$\koffp/1\,\mathrm{\frac{nm}{s}}$} -- (4,0)node[right]{$\konp/1\,\mathrm{\frac{mm}{s}}$};
\fill[green] (-1.25,-.08) rectangle (1.25,.08);
\fill[green] (-3.25,-.08) rectangle (-2.75,.08);
\fill[red] (.5,-.08) rectangle (1.25,.08);
\fill[orange] (.25,-.08) rectangle (.5,.08);
\fill[red] (-.5,-.08) rectangle (-1.25,.08);
\fill[orange] (-.25,-.08) rectangle (-.5,.08);
\fill[red] (-2.9,-.08) rectangle (-2.75,.08);
\fill[orange] (-2.9,-.08) rectangle (-2.95,.08);
\fill[red] (-3.15,-.08) rectangle(-3.25,.08);
\fill[orange] (-3.15,-.08) rectangle (-3.10,.08);
\foreach \x in {-3.5,-3.0,...,3.5}{
	\draw[] (\x,-.1) -- (\x,.1);
}
\draw[thick] (-2.5,-.15) -- (-2.5,.15);
\end{scope}
\begin{scope}[shift={(0,2*\barsep)}]
\node[] at (-8,0) {\singletxt};
\draw[stealth-stealth] (-4,0) node[left]{$\koffq/1\,\mathrm{\frac{nm}{s}}$} -- (4,0)node[right]{$\konq/1\,\mathrm{\frac{mm}{s}}$};
\fill[green] (-1.25,-.08) rectangle (1.25,.08);
\fill[green] (-3.25,-.08) rectangle (-2.75,.08);
\foreach \x in {-3.5,-3.0,...,3.5}{
	\draw[] (\x,-.1) -- (\x,.1);
}
\draw[thick] (-2.5,-.15) -- (-2.5,.15);
\end{scope}
\begin{scope}[shift={(0,3*\barsep)}]
\node[] at (-8,0) {\xbartxt};
\draw[stealth-stealth] (-4,0) node[left]{$\koffq/1\,\mathrm{\frac{nm}{s}}$} -- (4,0)node[right]{$\konq/1\,\mathrm{\frac{mm}{s}}$};
\fill[green] (-1.25,-.08) rectangle (1.25,.08);
\fill[green] (-3.25,-.08) rectangle (-2.75,.08);
\fill[red] (.5,-.08) rectangle (1.25,.08);
\fill[orange] (.25,-.08) rectangle (.5,.08);
\fill[red] (-.5,-.08) rectangle (-1.25,.08);
\fill[orange] (-.25,-.08) rectangle (-.5,.08);
\fill[red] (-3.225,-.08) rectangle (-3.175,.08);
\fill[orange] (-3.25,-.08) rectangle (-3.225,.08)
(-3.15,-.08) rectangle (-3.175,.08);
\fill[red] (-2.825,-.08) rectangle (-2.775,.08);
\fill[orange] (-2.775,-.08) rectangle (-2.75,.08)
(-2.85,-.08) rectangle (-2.825,.08);
\foreach \x in {-3.5,-3.0,...,3.5}{
	\draw[] (\x,-.1) -- (\x,.1);
}
\draw[thick] (-2.5,-.15) -- (-2.5,.15);
\end{scope}
\begin{scope}[shift={(\groupxsep,0)}]
\draw[stealth-stealth] (-4,0) node[left]{$\vonp/1\,\mathrm{V}$} -- (4,0)node[right]{$\voffp/1\,\mathrm{mV}$};
\fill[red] (-3.675,-.08) rectangle (-.25,.08);
\fill[orange] (-2.7,-.08) rectangle (-1.3,.08);
\fill[green] (-2.35,-.08) rectangle (-1.65,.08);
\fill[green] (2.0,-.08) rectangle (3,.08);
\draw (-2,.1) -- (-2,.7) node[above]{$\von$};
\draw (2.5,.1) -- (2.5,.7) node[above]{$\voff$};
\draw[thick] (1.5,-.15) -- (1.5,.15);
\foreach \x/\t in {-3.5/-1.0,-3.0/,-2.5/-0.8,-2/,-1.5/-0.6,-1/,-.5/-0.4,0/,.5/-0.2,1/,1.5/0.0,2/,2.5/10,3/,3.5/20.0}{
	\draw[] (\x,-.1) -- (\x,.1) node[above,fill=white,inner sep=1pt] {\t};
}

\end{scope}
\begin{scope}[shift={(\groupxsep,\barsep)}]
\draw[stealth-stealth] (-4,0) node[left]{$\vonp/1\,\mathrm{V}$} -- (4,0)node[right]{$\voffp/1\,\mathrm{mV}$};
\fill[red] (-3.675,-.08) rectangle (-.25,.08);
\fill[orange] (-2.35,-.08) rectangle (-1.65,.08);
\fill[green] (-2-.06,-.08) rectangle (-2+.06,.08);
\fill[green] (2.0,-.08) rectangle (3,.08);
\fill[orange](-2+.06,-.08)rectangle(-1.65,.08);
\fill[red] (2.4,-.08) rectangle(2.45,.08);
\fill[orange] (2.45,-.08) rectangle (2.48,.08)
(2.4,-.08) rectangle(2.37,.08);

\foreach \x in {-3.5,-3.0,...,3.5}{
	\draw[] (\x,-.1) -- (\x,.1);
}
\draw[thick] (1.5,-.15) -- (1.5,.15);
\end{scope}
\begin{scope}[shift={(\groupxsep,2*\barsep)}]
\draw[stealth-stealth] (-4,0) node[left]{$\vonq/1\,\mathrm{V}$} -- (4,0)node[right]{$\voffq/1\,\mathrm{mV}$};
\fill[red] (-3.675,-.08) rectangle (-.25,.08);
\fill[orange] (-2.35,-.08) rectangle (-2.0,.08);
\fill[green] (-2.0-.03,-.08) rectangle (-.25,.08);
\fill[green] (2.0,-.08) rectangle (3,.08);
\fill[red] (2.9,-.08) rectangle (3,.08);
\foreach \x in {-3.5,-3.0,...,3.5}{
	\draw[] (\x,-.1) -- (\x,.1);
}
\draw[thick] (1.5,-.15) -- (1.5,.15);
\end{scope}
\begin{scope}[shift={(\groupxsep,3*\barsep)}]
\draw[stealth-stealth] (-4,0) node[left]{$\vonq/1\,\mathrm{V}$} -- (4,0)node[right]{$\voffq/1\,\mathrm{mV}$};
\fill[red] (-3.675,-.08) rectangle (-.25,.08);
\fill[orange] (-2.35,-.08) rectangle (-2.0,.08);
\fill[green] (-2.0-.06,-.08) rectangle (-.25,.08);
\fill[green] (2.0,-.08) rectangle (3,.08);
\fill[orange](-1.3-.03,-.08) rectangle(-2+.06,.08);
\fill[red] (-1.65-.03,-.08) rectangle (-.25,.08); 
\foreach \x in {-3.5,-3.0,...,3.5}{
	\draw[] (\x,-.1) -- (\x,.1);
}
\draw[thick] (1.5,-.15) -- (1.5,.15);
\end{scope}

\begin{scope}[shift={(0,\groupysep)}]
\node[] at (-8,0) {\singletxt};
\draw[-stealth] (-4,0) -- (4,0)node[right]{$\Roffp/1\,\mathrm{k\Omega}$};
\fill[green] (-3,-.08) rectangle (2,.08);
\draw(-.5,.1) --(-.5,.7) node[above]{$\Roff$};
\foreach \x/\t in {-3.5/400,-3.0/,-2.5/600,-2/,-1.5/800,-1/,-.5/1000,0/,.5/1200,1/,1.5/1400,2/,2.5/1600,3/,3.5/}{
	\draw[] (\x,-.1) -- (\x,.1) node[above,fill=white,inner sep=1pt] {\t};
}
\end{scope}
\begin{scope}[shift={(0,\groupysep+\barsep)}]
\node[] at (-8,0) {\xbartxt};
\draw[-stealth] (-4,0)-- (4,0)node[right]{$\Roffp/1\,\mathrm{k\Omega}$};
\fill[green] (-2.5,-.08) rectangle (2,.08);
\fill[orange] (-2.5,-.08) rectangle (-2.75,.08);
\fill[red] (-2.75,-.08) rectangle (-3,.08);
\fill[red] (1.75,-.08) rectangle (2,.08);
\fill[orange] (1.5,-.08) rectangle (1.75,.08);
\foreach \x in {-3.5,-3.0,...,3.5}{
	\draw[] (\x,-.1) -- (\x,.1);
}
\end{scope}
\begin{scope}[shift={(\groupxsep,\groupysep)}]
\draw[-stealth] (-4,0)  -- (4,0)node[right]{$\Ronp/1\,\mathrm{k\Omega}$};
\fill[green] (-2,-.08) rectangle (3,.08);
\draw (.5,.1) -- (.5,.7) node[above]{$\Ron$};
\foreach \x/\t in {-3.5/2,-3.0/,-2.5/4,-2/,-1.5/6,-1/,-.5/8,0/,.5/10,1/,1.5/12,2/,2.5/14,3/,3.5/16}{
	\draw[] (\x,-.1) -- (\x,.1) node[above,fill=white,inner sep=1pt] {\t};
}
\end{scope}
\begin{scope}[shift={(\groupxsep,\groupysep+\barsep)}]
\draw[-stealth] (-4,0) -- (4,0)node[right]{$\Ronp/1\,\mathrm{k\Omega}$};
\fill[green] (-2,-.08) rectangle (3,.08);
\foreach \x in {-3.5,-3.0,...,3.5}{
	\draw[] (\x,-.1) -- (\x,.1);
}
\end{scope}
\begin{scope}[shift={(0,\groupysep+2*\barsep)}]
\node[] at (-8,0) {\singletxt};
\draw[-stealth] (-4,0) -- (4,0)node[right]{$\Roffq/1\,\mathrm{k\Omega}$};
\fill[green] (-1,-.08) rectangle (2,.08);
\fill[orange] (-1,-.08) rectangle (-1.5,.08);
\fill[red] (-1.5,-.08) rectangle (-3,.08);
\foreach \x in {-3.5,-3.0,...,3.5}{
	\draw[] (\x,-.1) -- (\x,.1);
}
\end{scope}
\begin{scope}[shift={(0,\groupysep+3*\barsep)}]
\node[] at (-8,0) {\xbartxt};
\draw[-stealth] (-4,0)-- (4,0)node[right]{$\Roffq/1\,\mathrm{k\Omega}$};
\fill[green] (-2,-.08) rectangle (2,.08);
\fill[orange] (-2,-.08) rectangle (-2.5,.08);
\fill[red] (-2.5,-.08) rectangle (-3,.08);
\fill[red] (1.75,-.08) rectangle (2,.08);
\fill[orange] (1.5,-.08) rectangle (1.75,.08);
\foreach \x in {-3.5,-3.0,...,3.5}{
	\draw[] (\x,-.1) -- (\x,.1);
}
\end{scope}
\begin{scope}[shift={(\groupxsep,\groupysep+2*\barsep)}]
\draw[-stealth] (-4,0)  -- (4,0)node[right]{$\Ronq/1\,\mathrm{k\Omega}$};
\fill[green] (-2,-.08) rectangle (3,.08);
\foreach \x in {-3.5,-3.0,...,3.5}{
	\draw[] (\x,-.1) -- (\x,.1);
}
\end{scope}
\begin{scope}[shift={(\groupxsep,\groupysep+3*\barsep)}]
\draw[-stealth] (-4,0) -- (4,0)node[right]{$\Ronq/1\,\mathrm{k\Omega}$};
\fill[green] (-2,-.08) rectangle (3,.08);
\foreach \x in {-3.5,-3.0,...,3.5}{
	\draw[] (\x,-.1) -- (\x,.1);
}
\end{scope}
\end{tikzpicture}

	\caption{Comparison of single gate and (combined) crossbar simulation results. A range of $\pm 50\%$ around the nominal value is plotted for each parameter. The results are color-coded: Green for correct \gls{imply} output, red for false output and orange for ranges inbetween, that are not covered by the simulation.}
	\label{fig:xbar_vs_single}
\end{figure*}

\subsection{Results analysis}\label{subsec:xbar_analysis}
In this section we compare the crossbar simulation results against the single gate results. 
As before, to be efficient, results are represented using our technique introduced in \Cref{subsec:postprocess}. \Cref{fig:xbar_full_results_variation_von_voff} shows a complete set of crossbar simulations for the parameters $\{\vonp,\voffp,\vonq,\voffq\}$.
\Cref{fig:xbar_vs_single} depicts the combined, i.e. worst case, results of all crossbar simulation setups explained in \Cref{subsec:xbar_method}, and the results of the single gate simulation, where the TTL threshold scheme was applied. The bars and color coding are identical to \Cref{fig:R_P_limit_m2_v,fig:R_P_limit_m2_R,fig:R_P_limit_m2_k,fig:R_P_limit_m0_v,fig:R_P_limit_m1_v}, \Cref{subsec:sim_analysis}. While the conclusions from \Cref{subsec:sim_analysis} remain true, unless noted otherwise, here we highlight the differences.

\subsubsection{Variation in voltage threshold}\label{subsubsec:xbar_v_var}
Given that the circuit is in a crossbar architecture, an increased number of errors due to threshold voltage variation can be expected in the crossbar simulation, when compared to the single gate simulation. Surprisingly, however, it is not significantly worse. \par 

There are three main differences: First, the initialization failure of $\voffq$ (initially shown in \Cref{fig:R_P_limit_m2_v}) does not arise in the crossbar simulation. However, there were initialization failures in the crossbar for $8\,\mathrm{mV}\leq\voffp\leq 9\,\mathrm{mV}$ ($-20\%$ to $-10\%$).  Having said that, as $\voff$ is of minor interest to the \gls{imply} operation, this can neither be considered an improvement, nor a degradation compared to the single gate. Second,  results indicate failures if both $\vonp$ and $\vonq$ are above  $-0.63\,\mathrm{V}\ (-10\%)$ at the same time. 
Based on the single gate simulation results (\Cref{subsec:sim_analysis}) and our recommendation to use \Cref{eqn:vonp_wmax} -- in combination with $R_Q=R_{Q,3}$ in \Cref{eqn:vpf} -- for device variability evaluation, this failure is predictable. As for the exact reason of this error, we assume that it is due to the increased state drift in $P$, as $|\vonp|$ is so low. In terms of operational range, the valid values for $\vonp$ and $\vonq$ are drastically restricted to the nominal value $\von$, as shown in \Cref{fig:xbar_vs_single}. It is only then that correct operations can be guaranteed.
However, if $\vonp < -0.63\,\mathrm{V}\ (-10\%)$ is ensured, a much greater range for $\vonq$ is admissible, similar to the case of the single gate in \Cref{subsec:sim_analysis}. Finally, the third difference is that the \gls{imply} operation fails for $\vonp<-0.77\,\mathrm{V}\ (+10\%)$ while $\vonq=\von$, as compared to $+20\%$ in the single gate simulation. Thus, the tolerable mismatch between $\vonp$ and $\vonq$ shrinks to $10\%$ within the crossbar. \par 

Apart from these differences the results of both simulations are identical, although \Cref{fig:xbar_vs_single} might not reveal it at the first look. This means that the proposed constraints for $\von$ and $\voff$ from \Cref{sec:math} can be applied to get a basic understanding of threshold voltage variability within crossbar architectures. \par 

\subsubsection{Variation in resistance limits}\label{subsubsec:xbar_r_var}
Varying the resistance limits of the memristors within the crossbar reveals some interesting results, as we can see in \Cref{fig:xbar_vs_single}. While \gls{imply} operations in the single gate simulation fail for $\Roffq\leq 800\,\mathrm{k\Omega}\ (-20\%)$, the crossbar simulation shows correct results down to $\Roffq=700\,\mathrm{k\Omega}\ (-30\%)$. We believe that this is due to the readout strategy applied to the crossbar, since the measured $R_Q$ after executing Case~3 (\Cref{tab:truth_table_imply}) is almost $1\,\mathrm{M\Omega}$ in a majority of the $-30\%$ simulation runs. Failures start occuring below $\Roffq\leq 600\,\mathrm{k\Omega}\ (-40\%)$. The range between $-30\%$ and $-40\%$ variation is not explicitly covered by our simulation steps. \par 

Furthermore, false \gls{imply} results within the crossbar come about at the upper and lower end of our simulated $\Roffq$ range, as well as at the upper and lower end of the simulated $\Roffp$ range. This is a combined effect, since those errors only occur if both, $\Roffp\leq 500\,\mathrm{k\Omega}\ (-50\%)$ and $\Roffq\geq 1.5\,\mathrm{M\Omega}\ (+50\%)$, or vice versa, are present at the same time. Interpreting this scenario based on the 1/2 or TTL logic thresholds from \Cref{tab:logic_thresholds}, one can see that if either $\Roffp$ or $\Roffq$ are below $500\,\mathrm{k\Omega}$, they are not interpreted as logic `0', but logic `1'. Thus, they do not fulfill Case~1 of the truth table, where $p=0$ and $q=0$ must be true. Applying the 1/3 logic threshold scheme, an off-resistance of $500\,\mathrm{k\Omega}$ yields an undefined logic state. Therefore, none of the cases in the truth table is fulfilled. Hence, such errors are predicted via logic thresholds alone and do not require further evaluation using the constraints defined in \Cref{sec:math}. \par 

Lastly, we should remark that simulation results for $\Ronp$ and $\Ronq$ in the crossbar are identical to the the single gate simulation results. 

\subsubsection{Variation in switching speed}\label{subsubsec:xbar_k_var}
Swichting speed variation does not pose a threat to single \gls{imply} gates, as deduced in \Cref{subsec:sim_analysis}. However, based on our simulation results (\Cref{fig:xbar_vs_single}), behavior within a crossbar is very different. For variations in $\konp$ and $\konq$ larger than $\pm 20\%$, the \gls{imply} operation fails. Further analysis of those failures reveals that it is the mismatch between $P$ and $Q$ which causes most errors.
If either $\Delta\konp\leq -20\%$ while $\Delta\konq\geq +20\%$, or $\Delta\konp\geq +20\%$ while $\Delta\konq\leq -20\%$, the operation result is wrong. This mismatch cannot be predicted by our constraints. Further, the simulation indicates failures for variation in $\koffp$ larger than $\pm 20\%$, as well as for $\Delta\koffq=\pm 40\%$. As $\koffp$ and $\koffq$ are never relevant during \gls{imply}, we infer that these are initialization errors. They can, however, be resolved by using a different initialization scheme than the one we applied. For example, using an additional readout cycle to confirm written initial states. Such a scheme provides feedback to resolve initialization errors before \gls{imply} is executed.

\section{Conclusion}\label{sec:conclusion}	
Device variability is one of the main challenges when implementing memristor-based logic. In this paper, we formulated novel constraints based on static switching conditions and state change dynamics. We note that the underlying causes of variation in device parameters are not differentiated by our methodology. Hence, environmental effects (such as temperature) causing parameter variation are taken into account by our constraints, just as process variations are. \gls{imply} operation results after a fixed timestep of execution were used as the metric to assess gate performance. In addition, different logic threshold schemes were considered. The derived constraints were put to the test in an extensive analysis for single gate and \xbarsize{} 1T1R crossbar and their simulation results were compared. An efficient simulation results presentation method was introduced and applied to find critical parameters. \par 

As a result of our analysis, variability in threshold voltages, especially $\vonq$, was identified as a major root of concern regarding correct operations. We conclude that the most dominant reasons for failure are predictable by our theoretical analysis for both the single gate and the crossbar. Therefore, our analysis and recommendations can be used for designing a reliable \gls{imply} gate. More specifically, we suggest to choose design parameters away from the borders of the recommended areas. Ideally, this distance should be chosen such that the typical (or maximum) variations, do not lead to crossing the borders of recommended area. Nonetheless, accompanying studies or simulations should be conducted to understand the non-deterministic errors, especially regarding voltage threshold- and switching speed mismatch within the crossbar, as well as state drift phenomena. \par 

Lastly, we note that our analysis can be used to decide whether a specific memristor technology and \gls{imply} logic are compatible. To that end, technology parameters need to be assessed based on the constraints for reliable \gls{imply} operations we extracted in this work. Further, considering technology-dependent parameter variation, an acceptable margin from the borders of the operating area must be ensured. Otherwise, chances for failures in \gls{imply} operations are increased. Hence, it would be better to use other technologies to implement the intended \gls{imply}-based circuits, or use other logics to implement the intended functionalities on the given technology.\par 

	\bibliographystyle{unsrt2authabbrvpp}    
	\bibliography{memristor_refs,references}

\begin{thebibliography}{10}

\bibitem{Niu2010}
D.~Niu \textit{et~al.}
\newblock Low-power dual-element memristor based memory design.
\newblock In {\em Proceedings of the 16th ACM/IEEE International Symposium on
  Low Power Electronics and Design}, ISLPED '10, pp. 25--30, New York, NY, USA,
  2010. ACM.

\bibitem{Ho2011}
Y.~{Ho} \textit{et~al.}
\newblock Dynamical properties and design analysis for nonvolatile memristor
  memories.
\newblock {\em IEEE Transactions on Circuits and Systems I: Regular Papers},
  58(4):724--736, April 2011.

\bibitem{Mohammad2013}
B.~{Mohammad} \textit{et~al.}
\newblock Robust hybrid memristor-{CMOS} memory: Modeling and design.
\newblock {\em IEEE Transactions on Very Large Scale Integration (VLSI)
  Systems}, 21(11):2069--2079, November 2013.

\bibitem{Baghel2015}
V.~S. {Baghel} and S.~{Akashe}.
\newblock Low power memristor based 7{T} {SRAM} using {MTCMOS} technique.
\newblock In {\em Fifth International Conference on Advanced Computing
  Communication Technologies}, pp. 222--226, February 2015.

\bibitem{Radakovits2019}
D.~Radakovits and N.~TaheriNejad.
\newblock Implementation and characterization of a memristive memory system.
\newblock In {\em 2019 IEEE 32nd Canadian Conference on Electrical and Computer
  Engineering (CCECE)}, pp. 1--5, May 2019.

\bibitem{zangeneh_design_of_1t1r_reram}
M.~Zangeneh and A.~Joshi.
\newblock {Design} {and} {Optimization} {of} {Nonvolatile} {Multibit} {1T1R}
  {Resistive} {RAM}.
\newblock {\em IEEE Transactions on Very Large Scale Integration (VLSI)
  Systems}, 22(8):1815--1828, Aug 2014.

\bibitem{kim2010cnna}
H.~Kim et~al.
\newblock Memristor-based multilevel memory.
\newblock In {\em CNNA2010}, pp. 1--6, Feb 2010.

\bibitem{Taherinejad2015ems}
N.~Taherinejad \textit{et~al.}
\newblock Memristors' potential for multi-bit storage and pattern learning.
\newblock In {\em EMS2015}, pp. 450--455, Oct 2015.

\bibitem{Taherinejad2016cce}
N.~Taherinejad \textit{et~al.}
\newblock Fully digital write-in scheme for multi-bit memristive storage.
\newblock In {\em CCE2016}, pp. 1--6. IEEE, 2016.

\bibitem{Pershin2012ieee}
Y.~Pershin and M.~Di~Ventra.
\newblock Neuromorphic, digital, and quantum computation with memory circuit
  elements.
\newblock {\em Proceedings of the IEEE}, 100(6):2071--2080, June 2012.

\bibitem{Thomas2013applied}
A.~Thomas.
\newblock Memristor-based neural networks.
\newblock {\em Journal of Physics D: Applied Physics}, 46(9):093001, 2013.

\bibitem{borghetti_memristive_stateful_logic}
J.~Borghetti \textit{et~al.}
\newblock ‘{Memristive}’ switches enable ‘stateful’ logic operations
  via material implication.
\newblock {\em Nature}, 464:873--876, April 2010.

\bibitem{talati_logic_design_magic}
N.~Talati \textit{et~al.}
\newblock {Logic} {Design} {Within} {Memristive} {Memories} {Using}
  {Memristor}-{Aided} {loGIC} ({MAGIC}).
\newblock {\em IEEE Transactions on Nanotechnology}, 15(4):635--650, July 2016.

\bibitem{CRS_proposal}
E.~Linn \textit{et~al.}
\newblock Complementary resistive switches for passive nanocrossbar memories.
\newblock {\em Nature Materials}, 9:403--406, May 2010.

\bibitem{Gupta2018}
S.~{Gupta} \textit{et~al.}
\newblock {FELIX}: Fast and energy-efficient logic in memory.
\newblock In {\em IEEE/ACM International Conference on Computer-Aided Design
  (ICCAD)}, pp. 1--7, November 2018.

\bibitem{Papandroulidakis2017}
G.~{Papandroulidakis} \textit{et~al.}
\newblock Crossbar-based memristive logic-in-memory architecture.
\newblock {\em IEEE Transactions on Nanotechnology}, 16(3):491--501, May 2017.

\bibitem{Rohani2017}
S.~G. Rohani and N.~TaheriNejad.
\newblock An improved algorithm for {IMPLY} logic based memristive full-adder.
\newblock In {\em 2017 IEEE 30th Canadian Conference on Electrical and Computer
  Engineering (CCECE)}, pp. 1--4, April 2017.

\bibitem{Taherinejad2019newcas}
N.~TaheriNejad \textit{et~al.}
\newblock A semi-serial topology for compact and fast {IMPLY}-based memristive
  full adders.
\newblock In {\em 2019 IEEE New Circuits and Systems symposium (NewCAS)}, pp.
  1--5, 2019.

\bibitem{guckert2018system}
L.~Guckert and E.~E. Swartzlander~Jr.
\newblock {\em System design with memristor technologies}.
\newblock Institution of Engineering \& Technology, 2018.

\bibitem{Radakovits2020tcasi}
D.~Radakovits \textit{et~al.}
\newblock A memristive multiplier using semi-serial imply-based adder.
\newblock {\em IEEE Transactions on Circuits and Systems I: Regular Papers},
  pp. 1--12, 2020.

\bibitem{CAS_physRealization}
N.~TaheriNejad and D.~Radakovits.
\newblock From behavioral design of memristive circuits and systems to physical
  implementations.
\newblock {\em IEEE Circuits and Systems Magazine}, 19(4):6--18, Fourthquarter
  2019.

\bibitem{waser_redox_reram_physics}
R.~Waser \textit{et~al.}
\newblock Redox-based resistive switching memories -- nanoionic mechanisms,
  prospects, and challenges.
\newblock {\em Advanced Materials}, 21:2632--2663, 2009.

\bibitem{menzel_switching_kinetics_ecm}
S.~Menzel \textit{et~al.}
\newblock Switching kinetics of electrochemical metallization memory cells.
\newblock {\em Physical Chemistry Chemical Physics}, 2013.
\newblock Cite this: Phys. Chem. Chem. Phys., 2013, 15, 6945.

\bibitem{menzel_reram_physics}
S.~Menzel \textit{et~al.}
\newblock Physics of the {Switching} {Kinetics} in {Resistive} {Memories}.
\newblock {\em Advanced Functional Materials}, 25:6306--6325, 2015.

\bibitem{cassuto_sneak_path_constraints}
Y.~Cassuto \textit{et~al.}
\newblock Sneak-{Path} {Constraints} in {Memristor} {Crossbar} {Arrays}.
\newblock {\em IEEE International}, pp. 156--160, 2013.

\bibitem{sneak_paths_closed_form}
M.~A. Zidan \textit{et~al.}
\newblock Memristor multiport readout: A closed-form solution for sneak paths.
\newblock {\em IEEE Transactions on Nanote}, 13(2):274--282, March 2014.

\bibitem{kvatinsky_device_variations}
N.~Wald and S.~Kvatinsky.
\newblock {U}nderstanding the influence of device, circuit and environmental
  variations on real processing in memristive memory using {M}emristor {A}ided
  {L}ogic.
\newblock {\em Microelectronics Journal}, 86:22--33, February 2019.

\bibitem{xie_robustness_of_memristor_logic}
L.~{Xie} \textit{et~al.}
\newblock On the robustness of memristor based logic gates.
\newblock In {\em 2017 IEEE 20th International Symposium on Design and
  Diagnostics of Electronic Circuits Systems (DDECS)}, pp. 158--163, April
  2017.

\bibitem{chen_imply_ron_not_reached}
Q.~Chen \textit{et~al.}
\newblock {A} {L}ogic {C}ircuit {D}esign for {P}erfecting {M}emristor-{B}ased
  {M}aterial {I}mplication.
\newblock {\em IEEE Transactions on Computer-Aided Design of Integrated
  Circuits and Systems}, 36(2):279--284, February 2017.

\bibitem{kvatinsky_imply_logic_design}
S.~Kvatinsky \textit{et~al.}
\newblock Memristor-based imply logic design procedure.
\newblock {\em IEEE 29th International Conference on Computer Design (ICCD)},
  pp. 142--147, 2011.

\bibitem{wan2010edl}
H.~Wan et~al.
\newblock In situ observation of compliance-current overshoot and its effect on
  resistive switching.
\newblock {\em EDL2010}, 31(3):246--248, 2010.

\bibitem{li2018imw}
C.~Li et~al.
\newblock In-memory computing with memristor arrays.
\newblock In {\em IMW2018}, pp. 1--4. IEEE, 2018.

\bibitem{li2018nature}
C.~Li et~al.
\newblock Analogue signal and image processing with large memristor crossbars.
\newblock {\em Nature Electronics}, 1(1):52, 2018.

\bibitem{chen_crossbar_array_model}
A.~{Chen}.
\newblock A comprehensive crossbar array model with solutions for line
  resistance and nonlinear device characteristics.
\newblock {\em IEEE Transactions on Electron Devices}, 60(4):1318--1326, April
  2013.

\bibitem{shin_data_dependent_statistical_model_analysis}
S.~Shin \textit{et~al.}
\newblock Analysis of passive memristive devices array: Data-dependent
  statistical model and self-adaptable sense resistance for rrams.
\newblock {\em Proceedings of the IEEE}, 100(6):2021--2032, 2012.

\bibitem{shin_data_dependend_model}
S.~Shin \textit{et~al.}
\newblock Data-dependent statistical memory model for passive array of
  memristive devices.
\newblock {\em IEEE Transactions on Circuits and Systems II: Express Briefs},
  57(12):986--990, December 2010.

\bibitem{kvatinsky_team}
S.~Kvatinsky \textit{et~al.}
\newblock {T}{E}{A}{M}: {T}hr{E}shold {A}daptive {M}emristor {M}odel.
\newblock {\em IEEE Transactions on Circuits and Systems I: Regular Papers},
  60(1):211--221, January 2013.

\bibitem{kvatinsky_vteam}
S.~Kvatinsky \textit{et~al.}
\newblock {V}{T}{E}{A}{M}: {A} {G}eneral {M}odel for {V}oltage-{C}ontrolled
  {M}emristors.
\newblock {\em IEEE Transactions on Circuits and Systems II: Express Briefs},
  62(8):786--790, August 2015.

\bibitem{strachan_memristor_model}
J.~P. {Strachan} \textit{et~al.}
\newblock State dynamics and modeling of tantalum oxide memristors.
\newblock {\em IEEE Transactions on Electron Devices}, 60(7):2194--2202, July
  2013.

\bibitem{jiang_stanford_model}
Z.~{Jiang} \textit{et~al.}
\newblock A compact model for metal–oxide resistive random access memory with
  experiment verification.
\newblock {\em IEEE Transactions on Electron Devices}, 63(5):1884--1892, May
  2016.

\bibitem{VTEAM_Spice}
D.~Radakovits \textit{et~al.}
\newblock Second (v2.0) {LTSpice} implementation of {VTEAM}, September 2019.
\newblock
  https://www.ict.tuwien.ac.at/staff/taherinejad/projects/\\memristor/files/vteam2.asc,
  https://www.ict.tuwien.ac.at/staff/\\taherinejad/projects/memristor/files/vteam2.asy.

\bibitem{knowm_datasheet}
Knowm.
\newblock {\em Knowm Self Directed Channel Memristors}, October 2019.
\newblock Rev. 3.2, https://knowm.org/downloads/Knowm{\_}Memristors.pdf, Last
  accessed: 11 March 2020.

\bibitem{kvatinsky_imply_principles}
S.~Kvatinsky \textit{et~al.}
\newblock {M}emristor-{B}ased {M}aterial {I}mplication ({IMPLY}) {L}ogic:
  {D}esign {P}rinciples and {M}ethodologies.
\newblock {\em IEEE Transactions on Very Large Scale Integration (VLSI)
  Systems}, 22(10):2054--2066, October 2014.

\bibitem{semiparallel_imply_adder}
S.~G. Rohani \textit{et~al.}
\newblock A semiparallel full-adder in imply logic.
\newblock {\em IEEE Transactions on Very Large Scale Integration (VLSI)
  Systems}, pp. 1--5, 2019.

\bibitem{TI_ttl_thresholds}
{Texas~Instruments}.
\newblock {\em {Logic} {Guide}}, 2017.
\newblock http://www.ti.com/lit/sg/\\sdyu001ab/sdyu001ab.pdf, Last accessed: 11
  March 2020.

\end{thebibliography}
	\begin{IEEEbiography}[{\includegraphics[width=1in,height=1.25in,clip,keepaspectratio]{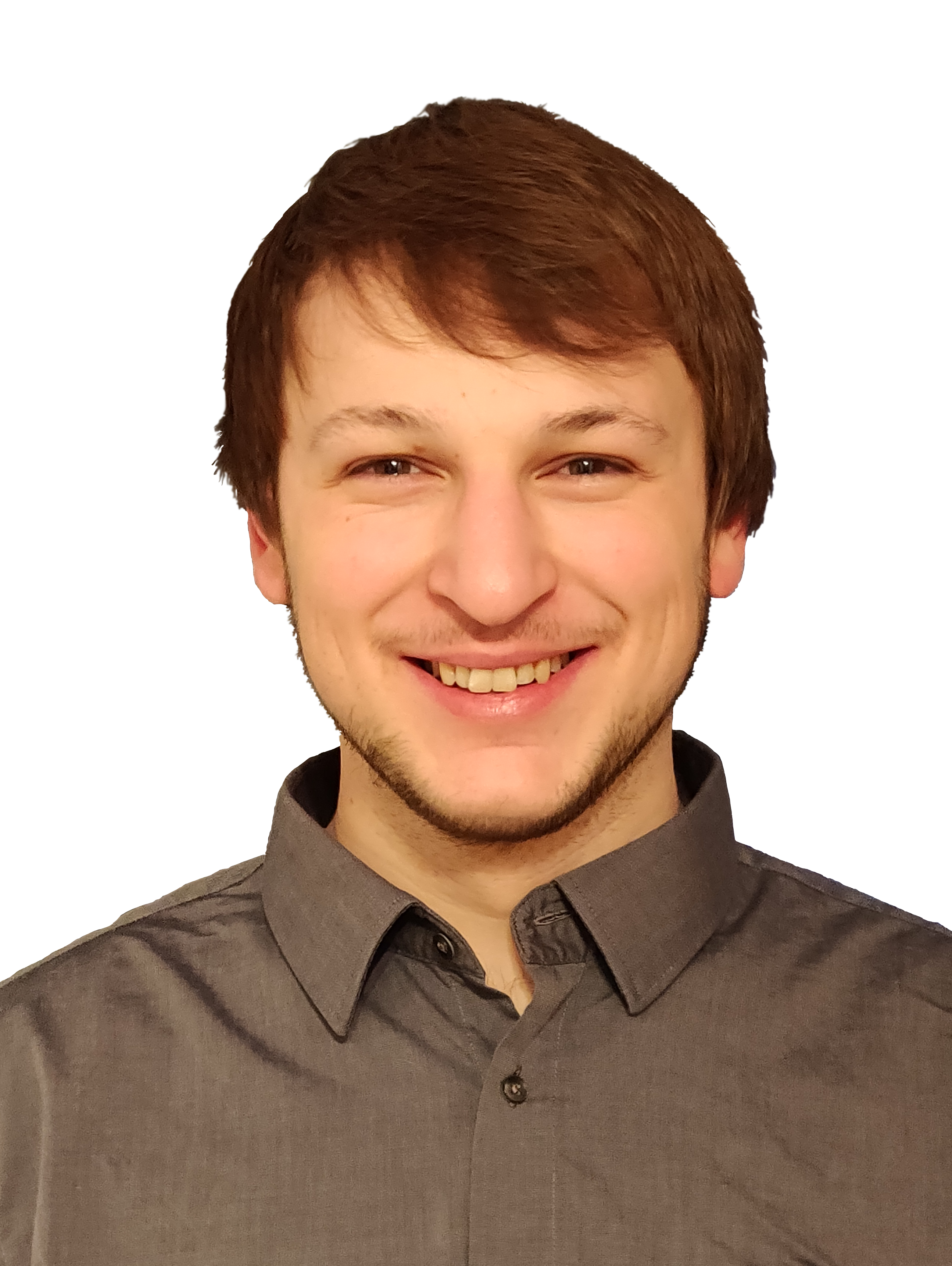}}]
		{Simon~Michael~Laube} is currently a B.Sc. student of electrical engineering and information technology 
		at the TU~Wien, 1040 Vienna, Austria. His B.Sc. thesis is on examining the robustness of 
		memristor-based material implication at the circuit/gate level.
	\end{IEEEbiography}	
	\begin{IEEEbiography}
[{\includegraphics[width=1in,height=1.25in,clip,keepaspectratio]{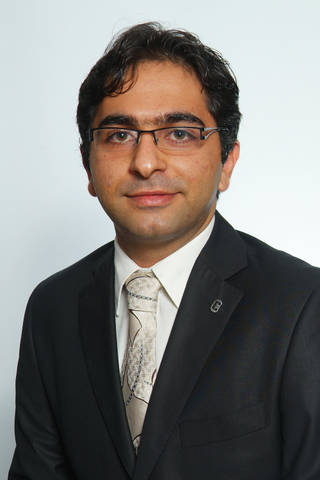}}]{Nima TaheriNejad} 
(S'08-M'15) received his Ph.D. degree in electrical and computer engineering from The University of British Columbia (UBC), Vancouver, Canada, in 2015.
He is currently a ``Universit\"{a}tsassistant'' at the TU Wien (formerly known as Vienna University of Technology as well), Vienna, Austria, where his areas of work include self-awareness in resource-constrained cyber-physical systems, embedded systems, in-memory computing, systems on chip, memristor-based circuit and systems, health-care, and robotics. He has published two books and more than 55 peer-reviewed articles. He has also served as a reviewer, an editor, an organizer, and the chair for various journals, conferences, and workshops.
Dr. Taherinejad has received several awards and scholarships from universities, conferences, and competitions he has attended.
In the field of memristive circuits and systems, his focus has been on physical implementations, reliability, memory, logic (particularly \gls{imply}), and in-memory computations.
\end{IEEEbiography}
\vfill
\newpage
\end{document}